\documentclass[12pt]{JHEP3}
\usepackage{amsfonts,amssymb}
\usepackage{amsmath}

\usepackage{epsf}

\def\T11{{T}^{1,1}}
\def\bear{\begin{eqnarray}}
\def\eear{\end{eqnarray}}

\newcommand{\vac}{{|0\rangle}}

\newcommand{\pa}{\partial}
\newcommand{\tr}{{\rm tr}}
\newcommand{\comment}[1]{}

\newcommand{\CO}{{\cal O}}
\newcommand{\pasl}{\pa\kern-.55em /}

\newcommand{\ksl}{k\kern-.55em /}

\newcommand{\ket}[1]{|#1\rangle}

\newcommand{\braket}[2]{\langle #1|#2\rangle}

\DeclareFixedFont{\xiiss}{OT1}{cmss}{m}{n}{12}
\DeclareFixedFont{\ixss}{OT1}{cmss}{m}{n}{9}
\DeclareFixedFont{\cmrnine}{OT1}{cmr}{m}{n}{9}
\newcommand{\field}[1]{\mathbb{#1}}

\newcommand{\BC}{{\field C}}
\newcommand{\BR}{{\field R}}

\newcommand{\CCs}{\hbox{\ixss C\kern-.4emI}}
\newcommand{\ZZs}{\hbox{\ixss Z\kern-.4emZ}}

\newcommand{\myfig}[3]{\begin{figure}[ht]
\begin{center}
\leavevmode
\epsfxsize=#2cm
\epsfbox{#1}
\end{center}
\caption{#3}
\label{fig:#1}
\end{figure}}

\title{ A Monte-Carlo study of the AdS/CFT correspondence: an exploration of
quantum gravity effects. 
 }
\author{David Berenstein, Randel Cotta\\
$^\dagger$ Department of Physics, UCSB, Santa Barbara, CA 93106}


\abstract{ In this paper we study the AdS/CFT correspondence for ${\cal N}=4$ SYM with gauge group $U(N)$, compactified on $S^3$ in four dimensions using Monte-Carlo techniques.
The simulation is based on a particular reduction of degrees of  freedom to commuting matrices
of constant fields, and in particular, we can write the wave functions of these degrees of freedom  exactly. The square of the wave function  is equivalent to a probability density for a 
Boltzman gas of interacting particles in six dimensions. From the simulation we can extract the density particle distribution for each wave function,
and this distribution can be interpreted as a special  geometric locus in the gravitational dual.
Studying the wave functions associated to half-BPS giant gravitons, we are able to show that the matrix model can measure the Planck scale directly. We also show that the output of our simulation seems to match various theoretical expectations in the large $N$ limit and that it captures $1/N$ effects as statistical fluctuations of the Boltzman gas with the expected scaling. Our results suggest that this is a very promising approach to explore quantum corrections and effects in gravitational physics on AdS spaces.
}

\begin{document}

\section{Introduction}

Many approaches to quantum gravity propose to replace the continuous geometry of spacetime by a discrete approximation whose underlying discreteness is concentrated at the Planck scale, the distance scale at which quantum effects are important. It is assumed that this discrete
theory is the correct description of gravitational physics at the Planck scale (recent examples of such ideas in concrete settings amenable to computer simulations can be found in \cite{RidS}).
The main technical problems with these approaches is to show that in the continuum limit (if it exists) one reproduces Einstein's theory of general relativity. This continuum limit would be related to quantization of gravity on large manifolds (as measured in Planck units), and classical geometry would be emergent in the sense that the continuum geometry is not part of the discrete problem, but that it only appears in a suitable limit.
 This will only be possible if one 
has a consistent quantum mechanical evolution (thereby solving the problem of time), as well as a semiclassical state of the theory that permits making the comparison to a semiclassical or classical approach based on perturbative techniques \footnote{To show that one can excite gravitational waves, or other quanta of the theory, one would need a family of such solutions, which are to be considered as small perturbations of the original solution.}.

Once one would obtain such a theory, it might be very hard to solve it in general and it is desirable that the discrete theory could be simulated on a computer. The main objective would be to try to analyze typical configurations and to ask what new phenomena can be understood in this way.
Clearly, if one wants to simulate an infinite space like Minkowski space or AdS geometries, one would need an infinitude of points and then such type of computational approach would need to be regulated somehow.

There are other approaches to a definition of quantum gravity that do not begin by assuming that spacetime is there as a fundamental object either. Within string theory, a very concrete proposal for a formulation of quantum gravity on AdS geometries that implicitly solves the problem of time 
is given by the AdS/CFT correspondence \cite{M}. The simplest example of the  AdS/CFT correspondence states that 
type IIB string theory on an asymptotically $AdS_5\times S^5$ spacetime with $N$ units of flux is exactly equivalent as a quantum mechanical system to the maximally supersymmetric SYM theory with gauge group $SU(N)$ in four dimensions, compactified on the conformal boundary of (global) $AdS_5\times S^5$ . This conformal boundary is $S^3\times \BR$. Thus, the quantum field theory is compactified on a round three sphere, and one has time translation invariance associated to the $\BR$ direction.  Since the SYM theory has a consistent quantum mechanical evolution and Hamiltonian, the problem of gravitational physics is to understand how the geometry
of the ten dimensional spacetime is encoded in the quantum system associated to the field theory. This can be rephrased as the following question: how does metric geometry emerge?

In particular, the SYM theory should in principle contain all of the (consistent) solutions of Einstein's equations with the required boundary conditions, so long as their radius of curvature is large in Planck units everywhere. These classical gravity solutions should be associated to some semi-classical state in the gravitational formulation, and for each of these one would have a dictionary to a particular state of the SYM theory (this is, a non-trivial wave function of the SYM theory). The dynamical evolution of these CFT states should reproduce in some way the dynamical evolution of the gravitational solution. 

In this paper we want to show that many\footnote{The word {\em many} in this phrase can be replaced by some or few, according to the detailed questions one wants to address.} aspects of this question can be analyzed numerically in the CFT, whereby one would recover a lot of information of various ten dimensional metrics from doing a computer simulation with finitely many degrees of freedom. The simulation we propose and execute in this particular paper is a Monte-Carlo algorithm that explores wave functions of an associated problem of $N$ identical particles in six dimensions. The relations of this particular system to the strong coupling limit of the CFT were explained and justified in \cite{B} and we will revisit them later on in the introduction. Given the wave functions, the distribution of particles in six dimensions in the thermodynamic limit (moderately large $N$) is concentrated on a five dimensional submanifold of $\BR^6$, and this submanifold is associated to a five-dimensional slice of the geometry of the ten dimensional spacetime.

In this paper we will be testing aspects of this idea with the simplest possible configurations, and we will show that there is an opportunity to study aspects of quantum gravity that are not 
accessible otherwise with these numerical methods. The main goal of this paper is to explain how such an identification is possible and to give numerical evidence supporting such an identification.

From a more general perspective, one of the biggest problems 
we have to face when trying to match gravity solutions and SYM states
is that the SYM states have to be evaluated at strong coupling, and very little is known about the behavior of SYM at strong coupling from first principles. In a similar vein, very little is known about the "most general solution of Einstein's equations" with the requisite boundary conditions. We need a simpler problem, just because our calculational abilities are limited.

Thus, it seems reasonable to restrict the problem further and exploit 
the supersymmetry of the supergravity and CFT problems to get to a more manageable set of states to compare.
In this sense, we can choose to find states that preserve many supersymmetries and symmetries.  With these restrictions it is possible to make some headway into the
problem of matching both sides of the duality. This is the main simplification of the problem that lets us reduce the degrees of freedom sufficiently to make a detailed comparison 
possible. In principle one could expect that other simplifications that use symmetries can result in a similar reductions of the degrees of freedom.

It seems reasonable to choose states that preserve $1/8$ of the supersymmetries.
The reason to choose such states, is that such a class of states can be associated to the chiral
ring of the SYM via the operator state correspondence, and the chiral ring has many non-renormalization theorems that permit a sensible comparison. This is also done to avoid the potential problem of having to evaluate a very complicated set of quantum corrections.
The other advantage of choosing such states, is that they are guaranteed to be associated to supergravity solutions of string theory, because the only one-particle states saturating the BPS bound in the gravitational theory are those associated to massless particles \cite{GKP,W}, and the only massless particles lie in the supergravity multiplet. Thus, in this case we do not have 
to test if we have a state made of heavy strings before comparing to geometry: one is guaranteed to be testing solutions of the type II supergravity theory in ten dimensions.

Our analysis of BPS states begins with the field theory (CFT) setup.
It is convenient to work in the canonical formalism of the field theory dynamics, for the field theory compactified on $S^3\times \BR$. 
The SYM theory multiplet contains as part of its spectrum 6 scalar fields $\phi^a(x,t)$ in the adjoint of $U(N)$, where $x$ is the position of the field in $S^3$ (the dynamics of the  $U(1)$ degrees of freedom are decoupled and can be ignored). 
To solve the theory in the free field limit, one would decompose the fields $\phi$ into spherical harmonics of the sphere $S^3$, in a normal mode decomposition
\begin{equation}
\phi^a(x,t) = \sum \tilde \phi^a_{l,m}(t) Y_{lm}(x)
\end{equation}
Here we are abusing notation as if the $S^3$ where a two-sphere. \footnote{In practice, the 
$S^3$ has an $SU(2)\times SU(2)$ isometry group, and the normal modes of a scalar appear in the $(n/2,n/2)$ representations of this group. Thus, the principal quantum number $n$
classifies the states with the same energy. In our abused notation, $l\sim n$ and $m$
includes the spin $z$ quantum numbers of both $SU(2)$ groups.}
The important point is that we have now a sum over a discrete set of oscillators, and it is straightforward to compute their energies, $w_{l,m} \sim l+1$.

A straightforward analysis of the chiral ring states shows that the only oscillators that are turned on near the free field limit are those with $l=0$ \cite{Btoy}, what one would call the $s$-wave of the normal mode expansion. These are just constant field configurations.

Because of these facts, and because the states we want to explore are supersymmetric, one
can assume that even at strong coupling only such constant field configurations are relevant, and that supersymmetry will guarantee that a semiclassical approach is essentially exact \cite{B}.

The advantage of this formulation for a partial solution via a BPS problem, is that it reduces the problem of infinitely many degrees of freedom of the field theory (all the spherical harmonics of the sphere) to finitely many degrees of freedom.

For these degrees of freedom, we can write an effective Hamiltonian, which is the dimensional reduction of the field theory on a sphere to the constant modes. This hamiltonian is
given by
\begin{equation}
H_{eff} =  \tr\left( \sum_{i=1}^6 \frac 12 (D_t X^i)^2 + \frac 1 2 (X^i)^2 +\sum_{a,b=1}^6\frac{g_{YM}^2}{8\pi^2} [X^a,X^b][X^a,X^b]\right)
\end{equation}
where the $X^i$ are six $N\times N$ hermitian matrices, and we have gauge transformations
$X^i \to U^{-1} X^i U$ where the $U$ are unitary matrices. The factors of $8\pi^2$ in the interactions arise from choosing canonical normalization of the matrix fields, once the volume of the three sphere is taken into account. A further reduction in degrees of freedom is obtained from going towards strong 't Hooft coupling, i.e. $g_{YM}^2 N$ large. This is because we usually expect that for random matrices, the eigenvalues are of order ${\sqrt N}$. Thus the term that is multiplied by $g_{YM}^2$ gets an extra factor of $N^3$, which tells us that at large $N$, the potential term dominates the dynamics.

The idea is that to determine the ground state, as well as the low lying excitations of the system we should look at the matrix configurations that
minimize the quartic term, and localize on those configurations. These configurations require the matrices $X^i$ to commute. There is still a family of matrix configurations that have these properties.

This is a very non-trivial statement, but there is strong evidence that this statement is correct, by comparing the energies of semi-classical string states \cite{HM} in $AdS_5\times S^5$ with the results of such an approach \cite{BCV} (see also \cite{hat}), where the string states are constructed by turning
on the off-diagonal modes of the $X$ fields in a perturbative expansion. These also confirm other perturbative resumations of planar diagrams \cite{SZ} and input from integrability structures \cite{Bei} by doing a calculation directly at strong coupling.

Being more careful about supersymmetry, one can notice that these configurations of commuting matrices  solve the F and D-term constraints of the field theory on flat space and are associated to the moduli space of vacua of the ${\cal N}=4 $ SYM.  Having commuting matrices is required in order to preserve supersymmetry, but it is not the only condition that one needs. One also needs a holomorphic wave function \cite{B}. Via a gauge transformation, we can choose all the  $X^i$ to be diagonal simultaneously for these configurations, and then the dynamics of supersymmetric states in the chiral ring is a particular quantization of the moduli space of vacua of ${\cal N}=4 $ SYM. This moduli space is described by $N$ (identical) particles on $\BC^3$ (or $\BR^6$). Because we also have a kinetic term and a quadratic term, we are going to obtain particular normalizable wave functions on this moduli space, and there will also be a preferred origin where $X=0$. The coordinates of these particles are the different (correlated) eigenvalues of the $X^i$ (this is in the same spirit as Matrix theory \cite{BFSS}).  Thus we can associate a six vector position per eigenvalue, as follows
\begin{equation}
\vec x_j = (X^1_{jj}, \dots, X^6_{jj})
\end{equation}

The quantum mechanics obtained in this way is almost free, as we have removed the explicit interactions between the matrices by requiring them to commute. If we diagonalize the matrices, we get an extra measure term $\mu^2$ from going to diagonal variables \cite{B}. This has been generalized  to some orbifolds
\cite{BC}, and suggests that one can generalize most of our techniques to the orbifold setup without too much trouble.

The problem of the study of the full CFT is reduced in this way to a study of a dynamical problem of $N$ particles in six dimensions with a given Hamiltonian. 
All we have to do now is solve the dynamics and find a large class of wave functions $\psi$ that solve the (time independent) Schrodinger equation. 
 Given any such wave function $\psi$, we then take a modified wave function $\hat \psi = \mu \psi$. The reason why it is convenient to absorb $\mu$ in the definition of $\psi$, is that $\mu^2$ will be the measure that is required for orthogonality of different eigenvalues of the Hamiltonian $H$.

From here $|\hat\psi|^2$ is a probability density in the usual sense, and it describes a gas of $N$ correlated particles on $\BC^3$. It is exactly the distributions of these particles in six dimensions that are interesting for our purposes. The reason for this, is that one is supposed to recover gravity in the classical limit as we send $N\to \infty$, and one expects the gravitational description to be some form of collective coordinates or thermodynamic description of the field theory 
problem. The simplest such collective coordinate is the density distribution of particles in this
six dimensional space of a "single particle" location.

It turns out that one can find a lot of approximate solutions of the Schrodinger equation for the Hamiltonian obtained by this procedure, and that these are all normalizable wave functions. The type of wave function that one obtains is similar to a thermal gas of particles with a confining potential and with logarithmic repulsions in six dimensions, which are induced by taking the logarithm of 
$\mu^2$. This type of problem is especially suited to Monte-Carlo simulations. This is very easy to implement on a computer. In this paper we will explain the algorithm we use to explore these configurations, as well as some of the improvements required to take $N$ large in a systematic way. We will also check our numerical results with the known theoretical results for this kind of thermodynamic gas in the large $N$ limit (unfortunately very little is known analytically).

The connection of this (commuting) matrix quantum mechanics model to gravity requires us to use the ideas of the AdS/CFT correspondence in detail in order to extract gravitational information from these configurations. We will explain our proposal for what kind of gravitational information we can expect to extract from these distributions of particles. By comparing with other known behavior of exact BPS solutions in gravity \cite{LLM}, we will be able to see that different topologies of spacetime are available to us.

Our paper is organized as follows. In section \ref{sec:MM} we explore the Hamiltonians that 
we are required to analyze, as well as the wave functions that we will study. We will also describe and analyze some properties of the these wave functions  in the thermodynamic limit
(large N limit). Some more results that are relevant for this section but are not required for the main argument are in appendix \ref{app:hw}. This section can be skipped on a first reading.
In section \ref{sec:AdS} we describe the chiral ring and it's relation to gravitational solutions. We discuss in detail the half-BPS states and we review briefly the work of Lin, Lunin and Maldacena \cite{LLM}. We use analogy to establish a relation between the geometry of the particle distributions obtained from the wave function calculations and the geometry of the dual gravitational solution. In section \ref{sec:mc} we describe in detail our Monte Carlo algorithm as well as how we propose to handle the large $N$ limit.
Our numerical results are compared to analytical large $N$ theoretical results in section 
\ref{sec:results}. We show that the numerical calculation agrees well with the theoretical 
expectation for the matrix models. We also study some configurations that are dual to LLM metrics, and that  are associated to different topologies of spacetime. We see a clear topology change. Moreover, we are able to show that the matrix model we are studying can be used to infer the size of the geometry in Planck units. This is, we are seeing evidence for effects that measure the Planck length in the simulation. On the other hand, some of the features of the simulation do not seem to agree with the LLM intuition. We end the paper with our conclusions and the prospects for future simulations along the lines we have outlined.

\section{Relevant exact results for quantum matrix models of commuting matrices}\label{sec:MM}

As described previously, we have reduced the problem of the full CFT
dynamics of ${\cal N}=4 $ SYM to a simple matrix quantum mechanics, and we further reduced that problem to a gauged quantum mechanical model of commuting matrices.

Because we can use gauge transformations to diagonalize matrices, the eigenvalues of the matrices are the only relevant degrees of freedom. However, permutations of the eigenvalues can be obtained by choosing to diagonalize the matrices via a different unitary operator. As such, the permutations of the eigenvalues are a gauge symmetry, which can be implemented by requiring that the wave function is invariant under such permutations. This gives us a problem of bosons
in a non-trivial setup.

Our purpose in this section is to explore the following Hamiltonian for $N$ identical (bosonic) particles in $d$ spatial dimensions
\begin{equation}
H= \sum_i -\frac 1{2\mu^2} \nabla_i \cdot \mu^2 \nabla_i + \frac 12 |\vec x_i|^2 \label{eq:HCommv2}
\end{equation}
where
\begin{equation}
\mu^2 = \prod_{i<j} |\vec x_i-\vec x_j|^2 \; .
\end{equation}
in the case where the dimension $d$ is even (and eventually it will be strictly greater than two).
This Hamiltonian was derived in \cite{B}. We refer the reader to that paper for details.

We should notice that if $\mu$ were replaced by one in \ref{eq:HCommv2} the Hamiltonian above would look just like that of N particles in a d-dimensional harmonic oscillator, this is
\begin{equation}
\tilde H_f \sim \sum_i -\frac 1{2} \nabla_i \cdot  \nabla_i + \frac 12 |\vec x_i|^2 \end{equation}

It is also clear that $\mu$ can only be important where $|\vec x_i -\vec x_j|$ is small.
It seems reasonable to try to solve the Hamiltonian with the Gaussian wave function that would solve the simple harmonic oscillator, plus some corrections where $\vec x_i-\vec x_j$ is small.

If we try 
\begin{equation}
\psi_0 \sim \exp( -\sum_i \vec x_i^2/2)
\end{equation} 
we can ask if $\psi_0$ is actually an eigenstate of the above Hamiltonian.

Direct manipulation shows that 
\begin{equation}
\nabla_i \psi_0 = \vec x_i \psi_0
\end{equation}

Similarly we can calculate
\begin{equation}
\nabla_i \cdot (\mu^2 x^i) \psi_0 = (\vec x^i)^2 \mu^2 \psi_0
+(\nabla_i \cdot x_i) \mu^2 \psi_0 +\psi_0\vec x_i \cdot \nabla_i \mu^2 
\end{equation}
from which we can calculate
$$
-\sum_i \frac 1{2\mu^2} \nabla_i \cdot \mu^2 \nabla_i\psi_0
$$
easily
 
We notice the following facts. The first term behaves like $\vec x_i^2 \psi_0$ and cancels the quadratic potential piece in the Hamiltonian. The second term is a constant times $\psi_0$.
(This constant is $N d$).
The third term is 
\begin{equation}
\psi_0\sum_i \vec x_i \cdot \nabla_i \mu^2
\end{equation}
We recognize that $\sum_i \vec x_i\cdot   \nabla_i$ is an Euler vector for dilatations where 
$\vec x_i\to \lambda \vec x_i$ simultaneoulsy. Clearly $\mu^2$ is an eigenvalue of this operator with eigenvalue $ N(N-1)$, because $\mu^2$ is a homogeneous function of the $x_i$ of this degree.

As such we obtain that 
\begin{equation}
H\psi_0 = E_0 \psi_0
\end{equation}
where $E_0= N(N-1)/2 + N d/2 $. It should be pointed out that in the case $d=1$, this is exactly the energy of $N^2$ harmonic oscillators, just as expected, because this is a single matrix model quantum mechanics. This was solved in \cite{BIPZ} (the review article \cite{Kleb} is also very useful to get some intuition of these setups).

It is convenient at this point to define a new wave function by a similarity transformation $\hat \psi = \mu \psi$. The measure of $\psi$ that makes the Hamiltonian self-adjoint is given exactly by $\mu^2$. This is, the probability for the particles to be located at various positions (within a fixed region $R$ about a particular set of positions) is
\begin{equation}
\int_R \mu^2 \psi^* \psi \sim \int_R \hat \psi^*\hat \psi
\end{equation}

Let us change the $d$ real coordinates of $\vec x_j$, $x_j^1 \dots x_j^d$ by the following complex combinations 
\begin{equation}
z_j^s = x_{j}^{2s-1}+ i x_{j}^{2s}
\end{equation}
by choosing a complex structure for the d dimensional space.

Now, let us consider a homogeneous polynomial of degree $m$ in the $z_j^s$ coordinates, and let us call this polynomial $P$. Furthermore, we need $P$ to be symmetric when we exchange the particles in $\BC^3$. With this additional information, we construct the wave function
\begin{equation}
\psi_P = \psi_0 P(z_j^s)
\end{equation}

The conjecture stated in \cite{B} is that all of these wave functions are the exact wave functions of the dual states to the chiral ring, and that the energy of the configuration is given exactly by the degree of the polynomial. In the appendix \ref{app:hw} we study these wave functions with the naive Hamiltonian written above where we have broken manifest supersymmetry by our choices of which degrees of freedom we are keeping. One can state that these wave functions are 
very good approximations to exact solutions of the non-supersymmetric Hamiltonian above. So if one were to do the same analysis being careful about supersymmetry, it is likely that one
might be able to prove exactness of the wave functions. It would be desirable to have such a formulation. For our purposes, all we need is the knowledge of the wave functions themselves.

Their $R$ charge (quantum numbers under rotations
of the $X$ fields) of the wave function is also the degree.

Notice also that when we calculate an overlap of two holomorphic wave functions with different degrees in the form
\begin{equation}
\int |\hat\psi_0|^2 P_m(z^s) \bar P_{m'}(\bar z^s) 
\end{equation}
we necessarily get zero. This is because the measure and $\hat \psi_0$ is invariant under phase rotations $z^s \to \exp(i\theta)z^s$ if we do  it on all the $z$ simultaneously, while
$P_m(z)\to \exp( im\theta)P_m(z)$.
This is related to a particular diagonal $SO(2)$ charge inside the $SO(6)$ R-charge of ${\cal N}=4 $ SYM theory, which defines an ${\cal N}=2 $ polarization.

Compared to the vacuum, these states have energy $m$, provided we redefine the zero of energy so that $E_0=0$. These states also have $R$-charge $m$. This is, the wave functions we have constructed have essentially the same energy and $R$-charge as expected for a BPS multiplet.
Because ${\cal N}= 4 $ SYM is conformally invariant, one can also use the operator-state correspondence. This correspondence states that for every local gauge invariant operator of the theory in (Euclidean) flat space ${\cal O}$, one can find a quantum state $\ket{\cal O}$ of the field theory compactified on $S^3$. The correspondence further states that the conformal dimension of ${\cal O}$ is the same thing as the energy of $\ket{\cal O}$ with respect to the ground state of the theory on $S^3$. The set of states we have obtained would have the same
$R$-charge and conformal dimension (energy) as elements of the chiral ring. Structurally, the chiral ring is also built by gauge invariant homolorphic operators, where the $F$ term constraints are imposed. These are holomorphic functions on the moduli space of vacua of the theory, and they would correspond exactly to symmetric holomorphic polynomials like $P$.

For the rest of the paper we will concentrate on studying the holomorphic wave functions described above and we will treat them as exact wave functions. It would be interesting if this
could be proved exactly by using supersymmetry arguments. 

\subsection{Thermodynamic behavior of the $N$-particle wave functions}

We now have a list of wave functions to analyze. They are all built by multiplying the ground state wave function $\hat \psi_0$ by a symmetric polynomial of the variables $z^i$.
We now want to find out what type of geometry these wave functions are associated to.

To begin, we want to study the ground state itself. We find that the square of the wave function, which has a probabilistic interpretation, takes the following form
\begin{equation}
|\hat\psi_0|^2 = \exp( -\sum_i \vec x_i^2 + \sum_{i<j} \log( |\vec x_i-\vec x_j|^2))
\end{equation}

Notice the similarity between this probability function and a Boltzman distribution
$\exp(-\beta \tilde H)$ for a gas of particles in $d$ dimensions, where
$\beta=1$ and $\tilde H = \sum_i \vec x_i^2 - \sum_{i<j} \log( |\vec x_i-\vec x_j|^2$, 
where the $\vec x_i$ are the positions of the particles. This is, we notice that we have
a gas of particles confined by a harmonic oscillator well, and that also have repulsive logarithmic interactions. If $d=2$ this is a Coulomb gas of particles (a two dimensional plasma) in a potential well. For higher dimensions this is a different problem.

If we are interested in a thermodynamic limit where $N\to \infty$ (meaning $N$ is taken to be very large), then we can hope that the gas will settle to a preferred thermodynamic configuration that will maximize the probability distribution, and that is well described by a density of particles in $d$ dimensions with some characteristic thermal fluctuations. 

We would approach this thermodynamic limit replacing sums by integrals, and introducing a density of particles $\rho(x)$. In this way the energy is given by 
\begin{equation}
\tilde H \sim \int (\vec x)^2 \rho(x) d^d x - \int\int d^d x d^d y
\rho(x)\rho(y)\log(|\vec x-\vec y|)
\end{equation}
subject to
\begin{equation}
\int \rho(x) d^d x = N
\end{equation}
and to $\rho(x) \geq 0$ \cite{B}.
 As is typical in thermal problems, we would first find the saddle point of $\tilde H$ that minimizes the energy. This was done  in detail in \cite{BCV}.  The main observation is that 
 we need to solve the following integral equation
 \begin{equation}
 x^2 +C = \int d^d y \rho(y) \log(|\vec x- \vec y|^2)
 \end{equation}
 in the region where $\rho$ has smooth support, and where $C$ is a lagrange multiplier enforcing the constraint. If $d$ is even, and greater than two, then 
 \begin{equation}
 (\nabla_x^2)^{d/2} \log(|\vec x-\vec y|) \sim \delta^d(\vec x -\vec y)
 \end{equation}
so that under the assumption of smooth support, taking derivatives with respect to $x$ and integrating over $y$ commute. If one uses this assumption, one applies $((\nabla_x)^2)^{d/2}$ on both sides of the equation. After this procedure one obtains the following equation 
\begin{equation}
0 = \int \rho(y) \delta^d(x-y) \sim \rho(x)
\end{equation}   
This contradicts that $\int \rho = N$. So one must conclude that the particle distribution $\rho$ has 
singular support in the thermodynamic saddle point limit.
 For the case above, it was found that due to spherical symmetry, one expects that the distribution is spherically symmetric and singular
$\rho(\vec x) \sim \delta(|\vec x| -r )$. The saddle of this ansatz occurs exactly when \begin{equation}
r= \sqrt{N/2},\label{eq:radius}
\end{equation} essentially independent of $d$ \cite{BCV}. We can treat this result as a guess for the answer. As we will see in section \ref{sec:results}, our simulation proves that this is correct in the thermodynamic limit.

For the case of ${\cal N}=4$ SYM, we need the special case $d=6$. In this case there are three (matrices of) complex variables that we need. It is customary to call them $X,Y,Z$. We have already used $X^a$ as real variables. The notation we will follow is that $X$ without an index represents $X^1+i X^2$.
There are other simple symmetric polynomials $P(X,Y,Z)$ one might consider other than one.

For example, take a single trace polynomial of $Z$, 
\begin{equation}
P_n = \sum_i (z^1_i)^n= \tr(Z^n)
\end{equation}
the wave function $\hat \psi_0 P_n$ is an allowed wave function. So is $\hat\psi_1= 
\hat\psi_0 (\tr (Z^n))^2= \hat\psi_0 P_n^2$ and 
$\hat \psi_2= \hat\psi_o \tr(Z^{2n})= \hat \psi_0 P_{2n}$. These two states 
are distinct functions of the eigenvalues of $Z$ with the same energy and $R$ charge, so they represent different states. What can we say about these two wave-functions? Here we need some more intuition. Since we are interested in comparing these wave functions with a gravitational dual configuration of $AdS$, we will get our intuition of what these objects represent from the expected dictionary of the AdS/CFT setup.

In the AdS/CFT dictionary established by Witten \cite{W}, each trace counts as a single graviton, so one expects that at large $N$, for fixed $n$, the two states $\hat\psi_1$ and $\hat \psi_2$ (properly normalized) are approximately orthogonal. This is, $\braket {\hat\psi_1}{\hat \psi_2} \sim \CO(N^{-1})$. The number of traces counts approximately the number of gravitons, and one can check that the traces do have an approximate oscillator algebra for a single matrix model, where $a_n^\dagger \ket \alpha \sim P_n(Z)\hat\psi_\alpha$. Here we have to assume this property, as we don't know how to calculate the norm of the corresponding states analytically.

Single graviton states on a given geometry do not correspond to classical states of geometry.
To get geometrical states, one would naturally expect that these are given by some type of coherent state. This is, our first guess for interesting geometric wave functions is to take an expression of the form $\exp (\sum_n t_n a^\dagger)\vac $ with finitely many $t_n$ different from zero.

However, we see that we have a problem with a naive extrapolation of Witten's result for this type of wave function. The reason is that
\begin{equation}
\exp(\sum_n t_n P_n(Z)) \hat \psi_0 
\end{equation} 
is not a normalizable state unless $t_n=0$ for all $n\geq 3$, because the trace dominates over $\hat\psi_0$ for very large values of the eigenvalues of $Z$, $z$. Thus we can not do quantum mechanics with such a state. What we need to fix this is good behavior at infinity, so that the wave function is $L^2$ integrable.

Let us  define $f(x) = \sum_n t_n x^n$. With this convention, $\sum_n t_n P_n(Z)= \tr f(Z)$.
To cure the bad behavior at infinity  we can require that $f$ have better behavior at infinity than any term in the series expansion of $f$. However, $f$ is a complex analytic function, so
to have such a property, $f$ has to be multivalued.

 The simplest behavior is for $f(x)$ to behave logarithmically at infinity. This is
 just thinking of a particular polynomial behavior of $\hat\psi/\hat\psi_0$, and we know that such wave functions are normalizable. The function $f$ will then have a branch cut in the complex plane. However $\psi$ can still be single valued, as it depends on $\exp(if)$. 
 Let us use instead
 \begin{equation}
 f \sim m \log( g(x))
 \end{equation}
 where $g$ is an arbitrary polynomial of $x$, and $m$ is a parameter that tunes the
 strength of the perturbation. For single valued wave functions, it should be an integer. For $x$ small, if $g(0)\neq 0$, we can expand $\log(g(x))$ in Taylor series, and we can approximate any polynomial $f$ to arbitrary 
 accuracy. Such a wave function would be given by
\begin{equation}
\hat \psi_0 \det (g(Z))^m
\end{equation} 

We expect that these wave functions are the ones that are important for geometry. Let us now consider a case where $g(0)\neq 0$, and where the first zero of $g$ happens for $x$ very large (much larger than 
$\sqrt {N}$, the typical radius of the sphere).

One would find then that the numerical value of $f$ is small for $x\sim\sqrt N$, and that $f$ is holomorphic in this region. As such, we can think of $f$ as a small perturbation of the confining potential, and that the saddle point of $\rho$ will react by a small change.

We can again go to the thermodynamic limit, and we find that we now need to satisfy the following integral equation
 \begin{equation}
 \vec x^2 +C -f(z)-\bar f(\bar z) = \int d^d y \rho(y) \log(|\vec x- \vec y|^2)
 \end{equation}
where $z= x^5+i x^6$. 
Thus $\nabla^2 (f(z) +\bar f(\bar z))=0$, because $f$ is holomorphic. Again, assuming $\rho$ is smooth leads to a contradiction. This leads us to
$\rho$ being given by a singular distribution as well.
The hard problem is to determine the distribution $\rho$ given $f$. For situations where $f$ is considered to be small, we expect that there is no change of topology, but that the geometric profile of the sphere gets deformed.

A more general analysis shows that
\begin{equation}
\hat \psi_0 \det(g(X,Y,Z))^m
\end{equation} 
will behave the same way for $g$ a polynomial in three complex variables, where the closest zero of $g$ to the origin is at distances much larger than $\sqrt N$. The factor of $m$ is an integer controlling the strength of the perturbation of the ground state geometry.

One can also consider situations where the zeroes of $g$ are more generic and let us say that these zeroes intersect the $S^5$ \footnote{Determinant operators are also related to giant gravitons in the AdS/CFT setup \cite{BBNS}. Thus one can identify the integer $m$ above with the number of $D$-branes in this case}. One finds that
the particles are repelled from the zeroes of $g$, because the wave functions vanish there.
Thus the support of $\rho$ does not intersect the zeroes of $g$, and the same arguments apply: $f$ is holomorphic (and multivalued) in the region of interest, but $f+\bar f$ is a real single valued function whose Laplacian vanishes. We expect in general that the description of all of these states corresponds to five-dimensional submaniolfds of $\BR^6$ of different topologies.

It is interesting to ask the following qualitative questions: what are the allowed topologies of these submanifolds (let us say for fixed degree of $g$)? Do these manifolds have boundaries, or lower dimensional pieces?  Are there universal features of topology transitions as we vary 
$g$ with fixed degree?

More quantitatively, we would like to know how the geometry of the submanifold correlates with the exact details of the wavefunction. Particularly, how the density of particles in the submanifold is related to other geometric properties of the embedding and if there is interesting scaling near topology transitions. One would also like to understand how finite $N$ effects blur the topology transitions. 

Our purpose in this paper is to show that these quantitative and qualitative questions about the particle distributions can be addressed numerically by simulating the wave functions using a Monte Carlo algorithm. 

Once these questions are understood (numerically) in the matrix model, one would like to find the gravitational dual description of these questions. Our proposal is that the geometric features discussed above correlate directly with the gravitational dual description and provide answers to quantum gravity questions for which there is no analytic understanding of finite $N$ effects. We will explain this correlation in the following section.

\section{AdS duals to chiral primaries}\label{sec:AdS}

Chiral primary operators by definition are BPS objects. Their main property is that their conformal dimension $\Delta$ is equal to their R-charge, $J$. These operators are related to BPS states for the field theory compactified on $S^3$. The superconformal algebra guarantees that $J$ and $H$ commute.  One can show that the twisted Hamiltonian $\tilde H= H-J$ is positive semi-definite in the free field limit and that only BPS states are annihilated by $\tilde H$. The classical solutions of $\tilde H=0$ are given exactly by constant configurations of the fields on $S^3$ and they are identical to the moduli space of vacua of the ${\cal N}=4 $ SYM theory \cite{B}.

On the gravitational side, one thinks of $J$ and $\Delta$ as particular isometries of $AdS_5\times S^5$ \cite{M}. For a point particle moving in this space $\Delta$ is the energy and $J$ is the angular momentum of the particle motion. Setting $\Delta=J$ tells us that the energy is equal to the momentum. This is, the particle is massless in ten dimensions. For type IIB string theory, all massless particles belong to the supergravity multiplet under supersymmetry transformations. Thus all BPS configurations with this amount of supersymmetry and with small quantum numbers should correspond to a gravitational excitation of the $AdS$ geometry. This is, one should be able to understand all these supersymmetric configurations in terms of supergravity. Because the different quanta respect the same supersymmetry, in principle it is possible that the gravitational attraction between two quanta is compensated exactly by exchange of the dilaton, giving rise to a non-linear superposition principle for solutions to Einstein's equations. The fact that the chiral primary operators have a ring structure (the chiral ring) tells us that is the expected behavior from the ${\cal N}=4$ supersymmetric field theory. Thus, it should be possible in 
principle to find fully non-linear solutions to the Einstein equations that respect the supersymmetries.

For large quantum numbers, one enters into the very non-linear regime of the classical gravity theory and it is possible to find new topologies and configurations that at first sight do not seem to be described by a gravitational background.

The simplest such configurations arise for BPS states with $J\sim N$, such that the angular momentum is only happening in an $SO(2)$ subgroup of the $SO(6)$ R-charge. These new semiclassical configurations correspond to giant gravitons that respect half of the supersymmetries. Giant gravitons are D3-brane configurations in $AdS_5\times S^5$ that respect some of the supersymmetries, in this case the same supersymmetries that ordinary gravitons
respect.  Giant gravitons were introduced in \cite{McGST} as the gravitational dual explanation for the fact that traces of $N\times N$ matrices $Z$, are not algebraically independent. Namely, that 
\begin{equation}
\tr(Z^{n+1}) \sim \sum \prod_i \tr(Z^k_i)
\end{equation}
with $k_i\leq N$.   We are also required to have $\sum k_i =N$. This means that the description of states as a Fock space in terms of traces is truncated. 
A more complete analysis shows that there are two types of giant gravitons. Some grow into 
the $S^5$ direction, and some grow into the $AdS$ direction \cite{GMT,HHI}. 

Their dual description in terms of field theory operators was conjectured in the works \cite{BBNS,CJR}. The giant gravitons growing in the $S^5$ are related to subdeterminant operators. These subdeterminants are the coefficients of the expansion of $det(Z+x)$. 
These are special cases of Schur functions related to completely antisymmetric representations of $U(N)$. The giant gravitons growing into the AdS were then conjectured to be related to Schur functions of completely symmetric representations of $U(N)$.

As we saw in the previous section, it was convenient to introduce determinants to build coherent states. Here we see that determinants are also interesting because of their relation to giant gravitons. We will comment more on this relationship further.

The description of \cite{CJR} was shown to be equivalent to an integer quantum hall droplet picture of two dimensional free fermions in the lowest Landau level with a confining quadratic potential \cite{Btoy}. In the fermion description, the two types of giants correspond to the holes and particles of the Fermi liquid. The giant gravitons corresponding to particles are naturally related to eigenvalues (this was also suggested in \cite{HHI}).The edge fluctuations are the ordinary gravitons. The particles and holes dissolve into edge fluctuations if their energy is of order $\sqrt N$. In this regime, the corresponding giant gravitons are very small, and all the relevant physics is captured by focusing on a plane wave limit  geometry. In this case this would be the maximally supersymmetric plane wave \cite{BFHP}. (See also \cite{BMN}).

It is natural to expect that when we place many mutually supersymmetric D3-branes moving on $AdS_5\times S^5$ on top of each other, that we will be able to replace the brane stack by a near horizon geometry free of singularities (these are non-dilatonic branes \cite{HS}, and the $AdS_5\times S^5$ global geometry of \cite{M} is a typical such example). In the half-BPS case, in the fermion language, placing the fermions as near to each other as possible, but away from the edge, produces new topologies of the 
droplet. The same is true if we work with many holes on top of each other. It is also natural to expect that we can turn on coherent states of gravitational perturbations of a system. These coherent states are edge waves of finite amplitude. Thus one expects that any macroscopic shape of droplets corresponds to a gravitational configuration.

This possibility of a simple fermionic description in field theory motivated Lin, Lunin and Maldacena to try to classify all solutions of supergravity that respect half of the supersymmetries of $AdS_5\times S^5$ and that have the same isometries as the
unbroken symmetries that the dual states preserve \cite{LLM}.

The metric needs to have the following symmetries: there is an unbroken $SO(4)$ because we chose spherically invariant configurations on the $S^3$ boundary. One has an unbroken $SO(4)$ subgroup of the R-charge (the little group of the highest weight state), because we only chose perturbations with $J$ being non-zero in an $SO(2)$ subgroup of $SO(6)$.

There is an additional unbroken translation symmetry because $\Delta=J$, so there is one extra generator of the superconformal group that annihilates the state and commutes with the two $SO(4)$ symmetries.

The natural ansatz for a space with this symmetry is of the form
\begin{equation}
ds^2 =  A(x) d\Omega_3^2+ B(x) d\tilde\Omega_3^2 - C(x)(d\tau+ V_i(x)dx^i)^2
+g_{ij} dx^i dx^j
\end{equation}
namely, a fibration with two three spheres $\Omega_3$ and $\tilde \Omega_3$, and a killing vector $\partial_\tau$. This leaves a three dimensional space of the $x$ variables where all the non-trivial dynamics is happening. After imposing the additional requirement of supersymmetry, the metric simplifies further, and one finds a preferred coordinate system for the remaining three variables, that is described by $x^1, x^2, y$. The full metric is then given by
\begin{equation}
ds^2 = - h^{-2}(dt +V_idx^i)^2+h^2(dy^2+dx^idx^i)+y e^Gd\Omega_3^2+y e^{-G}d\tilde \Omega_3^2
\label{eq:LLM}
\end{equation}
where $h^{-2}= 2y \cosh G$, and $V$ can be solved for if $G$ is known.
In the conventions of \cite{LLM}, one uses the auxiliary function
\begin{equation}
z= \frac 12 \tanh G
\end{equation}
and it was shown that $z$ satisfies a linear partial differential equation
\begin{equation}
\partial_i \partial_i z + y\partial_y \left(\frac{\partial_y z}{y}\right)
\end{equation}

There is a potential conical singularity in the metric at $y=0$. This is avoided if $z|_{y=0}$
takes the value $\pm 1/2$. One then finds that $h$ is finite and that the metric is regular. 
We impose these conditions as boundary conditions of $z$ at $y=0$. We see that the solution is determined by a two coloring of the $x^1, x^2$ plane (the regions where $z=\pm 1/2$, let us call them black and white). This is very similar to the two dimensional picture one obtains from the quantum hall droplet.

The interesting thing to notice here is that $y=0$ corresponds to a degeneration of one of the two three spheres to zero size. If one matches the supergravity solution to the quantum hall picture, the regions that have particles in them correspond to the locus where $\Omega_3$ (the sphere of $AdS_5$) has vanishing size (the black regions).

For the vacuum state, $AdS_5\times S^5$, in global coordinates, we have that
\begin{equation}
ds^2 = -\cosh^2(\rho) dt^2+d\rho^2 +\sinh^2\rho d\Omega_3^2+d\Omega_5^2
\end{equation}
and the locus where the radius of the $\Omega_3$ shrinks to zero size 
  is exactly the $S^5$ at the origin of $AdS_5$, namely $\rho=0$.  
  
Going back to the quantum hole analogy, if one considers a probe giant graviton of the particle type, 
one adds a small black disc outside the black region. In this disc the same $S^3$ that shrinks at the bottom of $AdS$, shrinks to zero size. Thus we can identify the locations of the $AdS$ giant gravitons with the regions where
the $S^3$ of the conformal boundary shrinks to zero size. For the other giant gravitons (the holes in the quantum hall picture), it is the other $S^3$ that degenerates to zero size.

A more general metric for the $1/8$ BPS states would have lower symmetry. Using the AdS/CFT dictionary, the states we need in the field theory are constant on the three sphere, as we have discussed previously. This symmetry would correspond in supergravity to an isometry of the metric. Thus we would still preserve the $SO(4)$ isometry, and this requires the metric to have a three sphere fibration. One would also have a Killing vector associated to the BPS bound.

This would give us
\begin{equation}
ds^2 = -A^2 (dt+V_idx^i)^2 + B d\Omega_3^2 + g_{ij} dx^i dx^j \label{eq:BPSans}
\end{equation}

This reduces the full gravitational problem to a system of non-linear partial differential equations in six dimensions. Such a possibility has been investigated in \cite{GMNO}, and in a slightly different context in \cite{Kim}.
One would expect that there are regions where this three-sphere fibration degenerates to spheres of zero size.  
These should be the locations where the giant gravitons that grow into $AdS$ are located.

\subsection{Proposal for comparison of matrix models and gravitational physics}\label{sec:physics}

So far, in the gravity side, we have noticed that the regions where the metric is interesting correspond to the locus where giant gravitons that grow into $AdS$ are located.
Fortunately, these giant gravitons can also be identified with the eigenvalues of the matrix model itself under the AdS/CFT dictionary \cite{HHI,Btoy,B}. 

Comparing to our previous analytical results of the CFT side in section \ref{sec:MM}, we noticed that for semiclassical states the eigenvalues of the matrices form well defined 
submanifolds in the six flat dimensions. It is natural to identify the singular eigenvalue distributions obtained from the field theory (in the large $N$ limit)  with the geometric locus where the size of the $S^3$ fiber of the $1/8$ BPS geometry vanishes. This $S^3$ survives in the conformal boundary of $AdS_5$, and it represents the $S^3$ on which the field theory has been compactified.

One also expects that there is locally a similar coordinate to $y$ in equation \ref{eq:LLM}, such that $y=0$ is the degeneration locus. In the case of the ground state, namely $AdS_5\times S^5$, the role of the $y$ coordinate is played by the radial direction of $AdS$, and the locus $y=0$ is the bottom of the $AdS$ potential well. 

In the eigenvalue picture, the $y$ coordinate can then be thought of as a transverse coordinate to the eigenvalue distribution in flat $\BR^6$. 

For the case of $AdS_5\times S^5$, the six dimensional space corresponds to the radial direction of $AdS$ and the $S^5$ together.  One can map this to the region outside the five-sphere distribution of eigenvalues. It is easy to convince oneself that exciting one eigenvalue in a BPS manner removes it from the five sphere and places it outside the $S^5$ distribution but not inside. 

Our proposal is that the support of the eigenvalue distribution represents exactly the degeneration locus of the three sphere in the full ten dimensional metric. As such, one can compare the degeneration locus of exact  supergravity solutions with their dual description.
The tests can be both qualitative, at the level of comparison of gross topological features, and they can also be more quantitative. For that, we need a proposal on how to extract metric data from an eigenvalue configuration.

The natural way to do this is to add small (but not massless) strings in a particular region. The length of the string serves as a probe of energy. It is also natural to add strings in the 
eigenvalue problem, similar to the work \cite{BCV}, by turning perturbatively the off-diagonal modes of the matrix model. This gives us, to first approximation, the induced flat space metric on 
the submanifold of the eigenvalue distribution.

One can in principle compare the energies of these two different descriptions of the same objects, in the degeneration locus. In the gravity setup, one has to take into account that time is warped to make the comparison. The comparison of metrics seems to give  
\begin{equation}
g^{Grav}_{ij}/A^2 \sim g^{Ind}_{ij}
\end{equation}

Our analysis suggests the following conjecture: for the $1/8$ BPS metrics with asymptotic $AdS_5\times S^5$ boundary conditions, the classifications of metrics can be
reduced to a six dimensional problem that is a base of an $S^3\times \BR$ fibration. This six dimensional (base) space should be diffeomorphic 
to an open connected subset of $\BR^6$ with a five-dimensional boundary. The region of $\BR^6$ that is carved out ends on the eigenvalue distribution in the dual field theory representation, and contains the region around infinity. The fact that the base is diffeomorphic to a subset of $\BR^6$ is a very strong topological constraint.

\section{The Monte Carlo algorithm}\label{sec:mc}

As we described in section \ref{sec:MM}, we are interested in computing the distribution of particles for an auxiliary statistical mechanical system of particles in $\BR^6$ (or in complex coordinates, particles in $\BC^3$) constructed as follows. We start with a wave function $\hat\psi$ of the form
\begin{equation}
\hat\psi
 =  \exp(-\sum_i \frac 12 \vec x_i^2+ \sum_{i<j} \log|\vec x_i -\vec x_j|) P(X,Y,Z)\end{equation}
where $P$ is a holomorphic function, invariant under permutations of the particles labeled by $i$. In particular, the ground state is the case for $P=1$. In general, for geometric states we expect that
$\log P(X,Y,Z) \sim m \sum_i \log g(x_i, y_i, z_i)$, where $m$ is an integer and $g$ is a polynomial in the complex coordinates with complex coefficients.

The wave function leads to a probability distribution  
\begin{eqnarray}
|\hat \psi|^2 &=& \exp(-\sum_i (\vec x_i^2-m [\log(g(x_i, y_i,z_i))-
\log(\bar g(\bar x_i, \bar y_i, \bar z_i))])+ \sum_{i<j} \log|\vec x_i -\vec x_j|^2)
\\&=& \exp(-\beta \tilde H)
\end{eqnarray}
that is equivalent to a Boltzman distribution for a gas at $\beta=1$ and with
\begin{equation}
\tilde H= -\sum_i (\vec x_i^2-m [\log(g(x_i, y_i,z_i))+
\log(\bar g(\bar x_i, \bar y_i, \bar z_i))]- \sum_{i<j} \log|\vec x_i -\vec x_j|^2
\end{equation}

The term $\sum_i (\vec x_i^2-m [\log(g(x_i, y_i,z_i))+
\log(\bar g(\bar x_i, \bar y_i, \bar z_i))]$ can be interpreted as a confining external potential, that is asymptotically quadratic. The term that sums  $\log|\vec x_i -\vec x_j|^2$ can be interpreted as repulsive logarithmic interactions between the particles. We are also required to study the system in the thermodynamic limit $N$ large, where $N$ is the number of particles in the system.

One expects on general grounds that the system will settle to some (thermal) equilibrium between the confining potential, the repulsion of the particles and with some extra thermal fluctuations. We want to study the (density) distribution of the particles in 
$\BR^6$, and we also want to measure the fluctuations around the equilibrium configurations. 

The equilibrium configuration is a typical configuration of the particles that dominates this thermal ensemble. For coarse grained observables (like the typical number of particles in some  sufficiently large specified region), a typical configuration is enough to characterize the ensemble accurately. Also, if one has a sufficiently symmetric situation, measuring on a single typical configuration can also let us measure thermal fluctuations of the ensemble by using symmetry operations to obtain more measurements out of a single distribution.

Our studies for this paper will only involve a special case of the set of distributions above, where $g(x_i, y_i, z_i)= x_i$, and $m$ is either zero (the ground state) or some other positive integer. 

For the case $m=0$, we expect that the distribution will have all particles forming a round five sphere shell, at a distance of 
\begin{equation}
r\sim \sqrt {N/2}
\end{equation} 
with some thermal fluctuations, and in principle $1/N$ corrections to the radius \cite{BCV}. The thermal fluctuations are also a 
$1/N$ effect. This is because in the large $N$ counting of t'Hooft, $1/N$ serves as a measure of $\hbar$, and our probability distribution arises from a quantum wave function. The quantum fluctuations of a wave function are typically an effect of order $\hbar$, and they translate to thermal fluctuations of our ensemble. In the gravitational theory, $1/N$ effects are
quantum gravity corrections to classical results.

Thus, in principle, we can measure some quantum gravity effects/corrections by looking at these fluctuations in the dual CFT description, that has been simplified to this thermal ensemble.

We use a standard Metrtopolis algorithm for a Markov process that navigates between the different configurations of particles in $\BR^6$. Between successive configurations $A,B$, we always accept the $B$ configuration if 
$\tilde H_B\leq \tilde H_A$, and if $\tilde H_B> \tilde H_A$, we generate a random number $s$ and compare $s$ to $\exp(-\beta(\tilde H_B-\tilde H_A))$. If $s$ is smaller than this quantity, then we accept the $B$ configuration. Otherwise we reject it.

Our algorithm to compute the thermal ensemble begins by setting up a random distribution of particles in six dimensions. We use the number of particles $N$ to set the size of this distribution. We disperse the particles randomly on a box of size $2 r$, where $r$ is the expected radius of the distribution, with each coordinate generated randomly in this range.

We generate new trial configurations by moving one particle at a time, $\vec x_i
\to \vec x_i +\delta \vec x_i$. The movement $\delta \vec x_i$ is generated randomly 
by varying the coordinates in a range $\pm\delta$ for each coordinate.
After moving one particle, we apply the metropolis criterion,
and we cycle between the particles in order.
After some number of iterations, we make $\delta$ smaller. This is designed to converge faster to a near equilibrium configuration, and then we use the smaller value to ensure that reasonable thermal fluctuations are generated somewhat accurately.
After some prescribed number of iterations $I$ we stop the calculations and look at the final configuration and consider it to be typical. We usually have half the iterations at one value of $\delta$, and then we make $\delta$ smaller by a factor of $5$.

 The typical values of $N$ that we use range between $100$ and 
$20000$. We found that moves with $\delta$ of order $2-5$ give good results. We used typically $300$ to $2000$ iterations per particle. Experimentally, we were finding convergence of results after about $100-150$ iterations.
Our computer code is a C program. We compiled our codes with gcc in various Apple computers.
We performed all our calculations with double precision, and we used the GNU scientific library random number generators to insure good statistical quality of the pseudo-random numbers.

The typical running time oscillated between 5 minutes and 48 hours depending on the number of particles and iterations, as well as on the speed of the computer processor that was used. The number of floating point operations that we perform scale roughly as $I N^2$, so making $N$ large is computationally expensive.

\subsection{Taking the large $N$ Limit}

We have found that taking $N$ large in the simulation directly is difficult (computationally expensive). If the configurations are going to be thermodynamic, it makes sense to coarse grain the dynamics in order to increase the effective value of $N$  without increasing the number of computations significantly. The simplest way to do that  is to assume that in the simulation each particle counts for $M$ particles near their vicinity, so that the effective $N$ is $N_{part}M$, where $M$ is a multiplicative factor associated to this coarse graining. This is taken care of by increasing the strength of the repulsive interactions by a factor of $M$. This is, we are simulating instead 
\begin{equation}
\exp(-\beta \tilde H) = 
\exp\left(-\sum_{i=1}^N (\vec x_i^2-m [\log(g)+\log(\bar g)])+ \sum_{i<j} M \log|\vec x_i -\vec x_j|^2\right)
\end{equation}
Roughly, particle $i$ feels the effect of $M$ particles at the location of particle $j$, but we are not averaging the position of particle $i$ over $M$ particles.
We need to verify that this is a good approximation for the questions we are asking. 

We do this by comparing distributions generated with the same $N_{eff}$, but with different
values of $M$ (the coarse graining factor). If we can show that numerically the answers are similar, then we can take $N_{eff}$ to be large by keeping the number of particles fixed and varying $M$.
We expect that this would induce errors of order $1/N$, where $N$ is the number of particles in the simulation, not $N_{eff}= NM$, because we are ignoring the $M-1$ particles in the vicinity of the $i-th$ particle that $i$ is not being affected by. So long as we can show that our errors are of this order, taking $N_{eff}$ to be very large is possible and we can compare to the thermodynamic limit.

\section{Results for sphere and Comparison to theory}\label{sec:results}

In the following tables, we show the numerical results for simulations. 
In particular 
we are able to 
show that in the large $N$ limit we approach the expected theoretical result from 
equation \ref{eq:radius}.
It is also evident that the effects of coarse-graining do not affect the calculations of the radius of the sphere substantially, and this allows us to extrapolate to relatively large values of $N_{eff}=NM$ by taking $M$ large.

Here we also present some results of varying the size of the motions of the particles in the Monte-Carlo algorithm. There is no substantial difference between runs where we take the particles and move them by bigger or smaller steps.

We can also plot a typical configuration of particles projects into the 1,2 plane, for 
a simulation with $N=1000$. The projection looks very round, and it is noticeably 
more dense in the center than in the outer edges, exactly as one would expect.

We can also show a histogram of the radius of each particle. It is evident by sight that the 
approximate radius of each particle is close to the expected value of $r\sim $ 22.36.

Our results comparing the simulation with coarse-graining to the full simulation 
for large numbers of particles are in the  table
\ref{tab:coar}. 

\begin{table}
\begin{center}
\begin{tabular}{|c|c|c|c|c|c|}\hline
N& M& $N_{eff}$ &        $r_{dist}$ &               $r_{th}$ &Relative difference\\
\hline
100& 1& 100& $7.204\pm 0.039$ & 7.07 & +1.9\%\\
\hline
200 & 1 &200& $10.094\pm 0.034$ & 10 & +.9\%\\
100 & 2 & & $10.068\pm 0.052$ & & +0.7\% \\
\hline
400 &1 &400& $14.212\pm0.031$&14.142 & +0.5\%\\
200&2& & $14.162\pm0.056$& &+0.1\% \\
100& 4& &$14.187\pm 0.0465$& &+0.3\%\\
\hline
800 & 1 & 800 & $20.038  \pm 0.014$ & 20 & +0.2\%\\
400 & 2 &        & $20.025\pm 0.021$ &        & +0.1\%\\
200& 4 &        & $20.0165\pm 0.023$&        &+0.1\%\\
100 & 8      &    & $19.928\pm 0.041$ &   & -0.4\% \\
\hline
1600& 1& 1600& $28.315\pm0.013$ & 28.284& +0.1\%\\
800& 2&    & $28.305\pm 0.012$ & & +0.1\%\\
400& 4 & & $28.301\pm0.028$& & +0.1\%\\
200& 8 & & $28.252\pm0.032$& &-0.1\% \\
\hline
\end{tabular}
\caption{Comparison of coarse graining to full simulation}\label{tab:coar}
\end{center}
\end{table}
In the table we calculate the radius of the distribution by using
\begin{equation}
r_{dist} = \sqrt{ \frac 1 N \sum_{i=1}^N \vec r_i^2}
\end{equation}
and we are showing the theoretical value on the right, $\sqrt{N_{eff}/2}$. We also show the
relative percent difference $(r_{dis}-r_{th})/r_{th}$ to the theoretical value.
The error bars are based on samples of $10$ configurations per distribution, where we quote the standard deviation of a sample around the mean value.
Here, it is useful to notice that just like in the case of spherical membranes in Matrix theory \cite{KT}, there are various possible definitions of the radius once the statistical fuzzyness of the sphere
is taken into account. The one we chose seems like a sensible definition, as it is given by
\begin{equation}
r^2 \sim \frac 1 N \tr(\vec X^2)
\end{equation}
the simplest  matrix model correlator that one could use.

Our error bars are the statistical errors
of our sample, but they do not include an estimate of systematic errors of the code, or show that correlations between different samples associated to the same distribution are absent.
Our purpose in this paper has been to establish that this type of simulation is a viable avenue to understand 
geometrical information using a very simplified matrix model analysis, and this requires us to take $N_{eff}$ as large as possible. 

As can be clearly seen, for the most part the effect of coarse graining is consistent with
a full simulation, and it tends to lower the value of $r_{dist}$ with respect to the full simulation.
This is expected, as in our coarse graining procedure we are ignoring the self repulsion of the eigenvalues that have been coarse grained into a single one. This is a systematic effect that is 
of order $M/N_{eff} = 1/N$ and tends to lower the size of the distribution (there is a slightly smaller net effective repulsion). We should also notice that the size of the error bars in the radius distribution are very similar between the coarse-grained distribution and the full simulation. Also, the results are close to the expected theoretical result. Looking at the simulations with $M=1$, we notice that as $N$ gets larger, the statistical error bars in the radius tend to go down in size. For $N=1600$, the relative error is $0.1\%$.

Table \ref{tab:larN} shows a simulation with $N=400$ where we just change the coarse-graining
multiplicity $M$ through various values. The relative difference is of order $0.1\%\sim 1/1000$
for large values of $N_{eff}$. This is smaller than  $1/400$ by a factor of $2$, a systematic $M/N_{eff}$ effect. We can estimate this effect by comparing
$\sqrt{N M/2}$ to $\sqrt{(N-1)M/2}$, the first is the effective radius one would get for $N_{eff}=NM$, while the second radius is the mean field 
theory (the particle $i$ feels the repulsion of $(N-1)M$ particles and the scaling of the problem. 
This shows a relative difference of order $- 1/2N \sim -1/800$, exactly like we are seeing in the data. Our results are therefore consistent with the theoretical large $N$ limit when we take this systematic effect into account.

\begin{table}\begin{center}
\begin{tabular}{|c| c| c| c| c|}\hline
M & $N_{eff}$ & $r_{dist}$ & $r_{th}$& Relative difference\\
\hline
1 & 400 & $14.213\pm 0.031$ & 14.142 & +0.5\% \\
2 & 800 & $20.025\pm0.021$ & 20& +0.1\% \\
4 & 1600& $28.301\pm 0.028$&28.284& +0.1\%\\
10 & 4000& $44.689\pm0.255$& 44.72& -0.1\% \\
20 & 8000 & $63.178\pm 0.027$& 89.44& -0.1\% \\
50 & 20000& $99.890\pm 0.012$& 100& -0.1\% \\
70 & 28000 & $118.176\pm 0.025$&118.32& -0.1\% \\
100&  40000 & $141.237\pm0.018$& 141.42& -0.1\% \\
1000& 400000 & $446.663\pm 0.027$&447.21& -0.1\% \\
\hline
\end{tabular}
\caption{Coarse grained distribution for large $N$ limit.}\label{tab:larN}
\end{center}
\end{table}

\subsection{Density fluctuations and $1/N$ counting}

As we have shown above, our numerical results seem to be converging to the radius of the particle distribution that is obtained from the saddle point approximation. 
It is convenient also to test the density fluctuations along the five-sphere, to test how spherically homogeneous it is. One would also like to understand how thin is the sphere, as in the saddle point limit we are suppose to be comparing it to a delta function distribution at fixed radius.

To see whether the system has rotational symmetry, one can for example take a collection of points generated this way and project them onto the $12$ plane. This can give us an idea of how spherically symmetric the distributions are.
We present here an example with $N=2000$ in figure \ref{fig:pro12.eps}.

\myfig{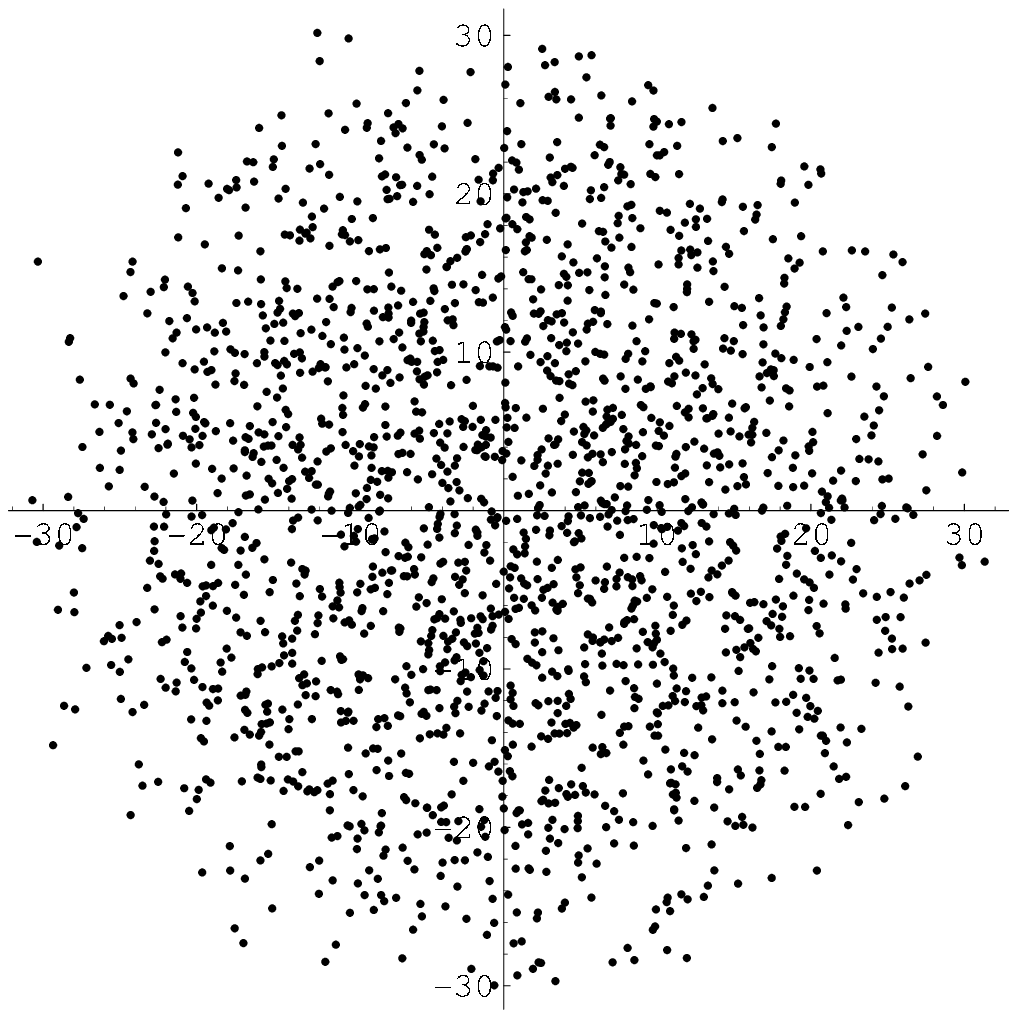}{7}{Projection onto the $12$ plane of a typical configuration of 2000 particles.} 

By inspection, the distribution looks fairly circular. We should also remember that this is projection of a 5-sphere embedded in flat six dimensions. This gives us an image of 
the 5-sphere as a 3-sphere fibration over a disc. The radius of the $S^3$ fiber is
$r_3^2= (r^2-r_{12}^2)$ in terms of the $(r_{12})$ radial coordinate, and $r$, the radius of the distribution in $6$ dimensions. The number of points 
for each sphere element scales with the radius of the 3-sphere, so the density of points
at the center has to be much higher than at the edges. This is observed in the figure. This makes it hard to estimate how spherical the object is, because our sight perceives mostly the edges of the disc, where the density of coverage is small.

To do better work, one would want to count the number of particles in cones of given angles. However, this can depend very much on how one chooses to slice these angular regions and how one triangulates the six sphere into smaller pieces. Instead, as is usual in other situations, one would much rather consider the multipole expansion of the distribution. Since the particles are all expected to be essentially at the same radius, the multipole expansion will give us directly the angular fluctuations of the density into different multipole moments.

Here, we will just calculate the quadrupole moment squared of the distribution. To ensure that we are doing things carefully, we can expand the square of the quadrupole moments on all 2-planes.

The typical quadrupole moment (along the 12 plane lets say) can be written as
$
Q \sim \Re e(\tr(Z^2))
$
where we choose $Z= X_1+iX^2$. We want to normalize the quadrupole to reflect just the angles and not the radius, so we should use 
\begin{equation}
Q = \Re e(\tr(Z^2/r^2))
\end{equation}

Because of spherical symmetry, the average quadrupole will be exactly zero. But to understand 
the typical fluctuation, we would want to calculate 
\begin{equation}
Q^2 \sim \langle \tr (\bar Z^2/r^2)\tr (Z^2/r^2)\rangle \label{eq:nquad}
\end{equation}

These are typical correlators in matrix models. In the large $N$ limit, formal developments around perturbative constructions 
lead to a $1/N^2$ expansion into non-planar diagrams, so long as one can think of the solution of the matrix model as a resummation of a perturbation series. We should notice that the Monte-Carlo simulation is non-perturbative in nature, so it is possible to observe different behavior than the expected perturbative result in the simulation.

 In our case, we have that the radius $r$ is of order $\sqrt N$, so  that a correlator 
 $\tr(Z\bar Z)$ scales like $N^2$, and $\tr(Z\bar Z)^2$ would scale like $ N^4$. If our results can be understood in this type of framework, we would expect that the numerical values of various objects will have a $1/N$ expansion. The 
tree level value of $\tr(Z^2)$ is zero, so for $|\tr(Z^2)|^2$ we expect a result which is suppressed with respect to $N^4$ by a factor of $N^2$. If we correct for the normalization of the radius, as in equation \ref{eq:nquad}, so that we can compare different droplets of varying $N$, we expect that these correlators are all of order one. This can be tested in our numerically generated distributions for various values of $N$. This is shown in table \ref{tab:quad}.

The $Q$ is generated by averaging over quadrupoles in various planes. For each  configuration of eigenvalues, there are 20 linearly independent quadrupole moments.

In practice we calculate the six numbers
\begin{equation}
S_{ij} = \tr (X_i^2 - X_j^2)/r^2
\end{equation}
and the fifteen numbers
\begin{equation}
Q_{ij} = \tr( 2 X_i X_j) = \tr [ \frac 12 \left((X_i+X_j)^2- (X_i - X_j)^2\right)]
\end{equation}
and then we average their squares, to obtain $Q^2$. These $21$ numbers contain all $20$ independent quadrupole moments of a single distribution of particles.
We then average these over ten different distributions with the same number of particles.

\begin{table}
\begin{center}
\begin{tabular}{|c|c|}  \hline  N & $Q^2$  \\
\hline
100 & 2.74\\
400 & 2.56\\
800 & 3.28\\
1200 & 2.82\\
1600 & 3.21\\
\hline
\end{tabular}
\caption{Quadrupole moment table} \label{tab:quad}
\end{center}
\end{table}

If instead we had a gas of uncorrelated particles at constant radius, the total quadrupole 
moment would scale with $N^{1/2}$ where $N$ is the number of particles. The matrix model behavior signals a highly correlated system compared to a free gas. In this sense, it can be considered as a type of liquid (the same gas model in two dimensions is the particle distribution associated to a quantum hall fluid).

Indeed, our table \ref{tab:quad} shows that the total quadrupole moments 
do not vary substantially between various values
of the number of particles and that the numerical values seem to agree with the expectations of large $N$ matrix model statistics. Also notice that any growth in the quadrupole moment with increasing $N$ is at best marginal. We also don't have error bars associated to these numbers $Q^2$: we don't have enough information to compute them. The table should be interpreted as providing numerical evidence for the expected large $N$ scaling of a matrix model.

It should be noticed that the multipole expansion in angular coordinates would give us correlators of the form
\begin{equation}
M^2\sim \langle \tr (\bar Z^n/r^n)\tr (Z^n/r^n)\rangle
\end{equation}
while some multipole correlation functions would be of he form
\begin{equation}
M^2\sim \langle \tr (\bar Z^m/r^m)\tr(\bar Z^n/r^n)\tr (Z^p/r^p)\rangle
\end{equation}
and in principle can be compared to three point functions in the conformal field theory, and via the AdS/CFT, to supergravity. These three point functions for half-BPS states were first studied in detail in \cite{LMRS}, and suggested various non-renormalization theorems. These have been 
analyzed in more detail in \cite{DFS}.

To do a useful numerical comparison we expect to need a large statistical sample, as these normalized correlators will be expected to be of order $1/N$. Such a statistical sample is beyond the scope of the present paper. This is currently under investigation \cite{BCH}.

Similar to the angular fluctuations  of the density described above, one can also slice the 
points radially and try to understand the radial distribution of particles in more detail.
In particular, one would like to determine if the radial distribution of a single particle
is approximately Gaussian or not. Naively, non gaussian behavior should be suppressed by 
some (possibly fractional) powers of $1/N$ and could correspond to some non-trivial quantum
gravity effect.

We find that the typical width of the radial distribution of particles is always of order $1$ ($0.86$ in the example), while the radius is of order $\sqrt N$. We also find no noticeable deviation from a gaussian distribution for the distribution of the radii for a simulation with $N= 2000$.  this is encoded in the figures \ref{fig:hist.eps} and \ref{fig:compare.eps}.

\myfig{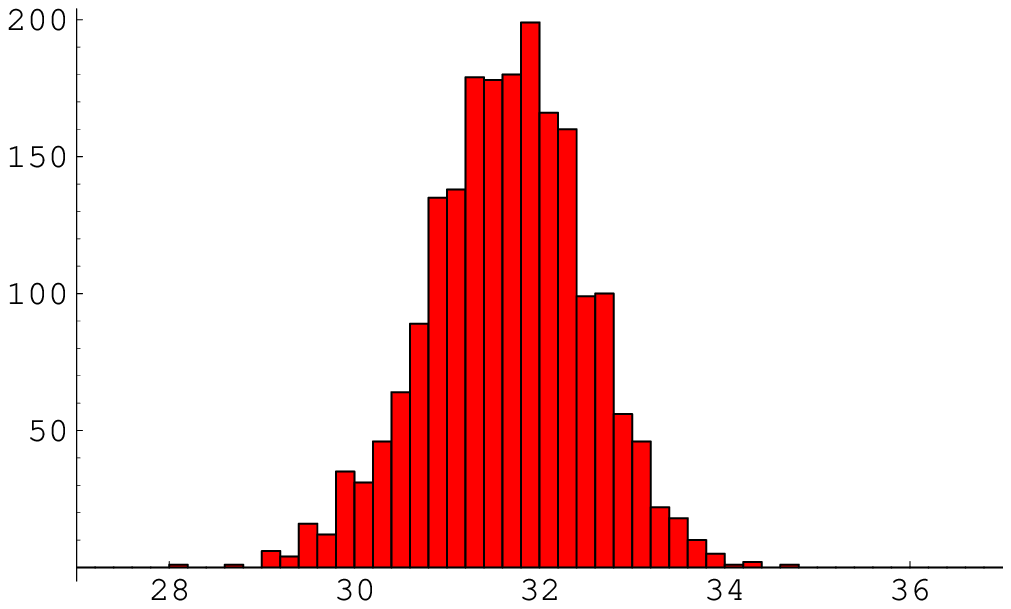}{5}{Binned radial data for one sample of $N=2000$ on 50 bins between $27$ and $37$, with an expected radius  $r\sim 32$}

\myfig{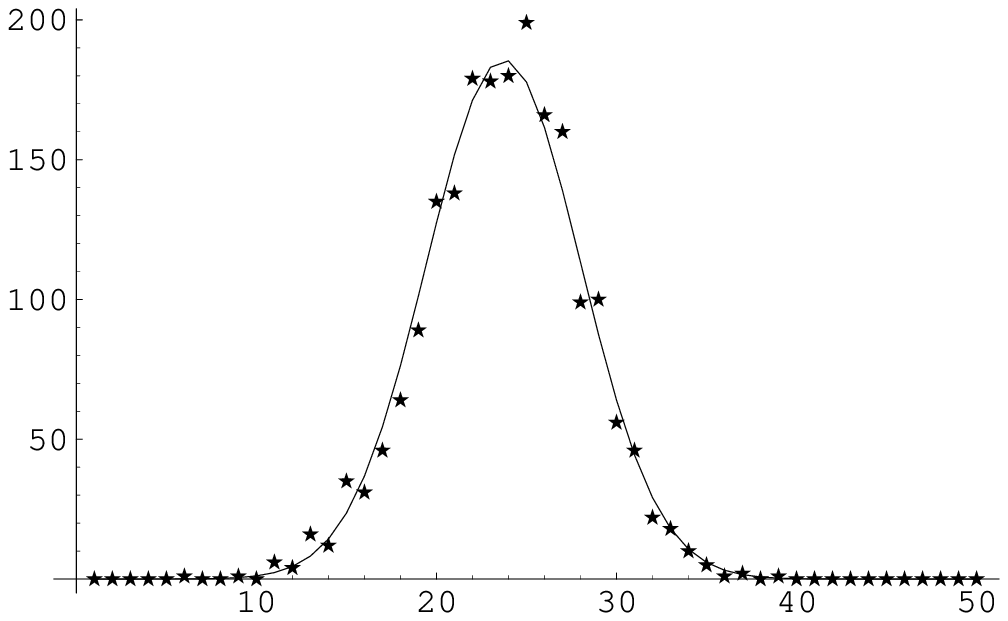}{ 5}{ Comparison between simulation and a Gaussian fit by bin.}

Finally, we can also look at the radii of the individual particles in the order they were 
generated. This is shown in figure \ref{fig:rad.eps}

\myfig{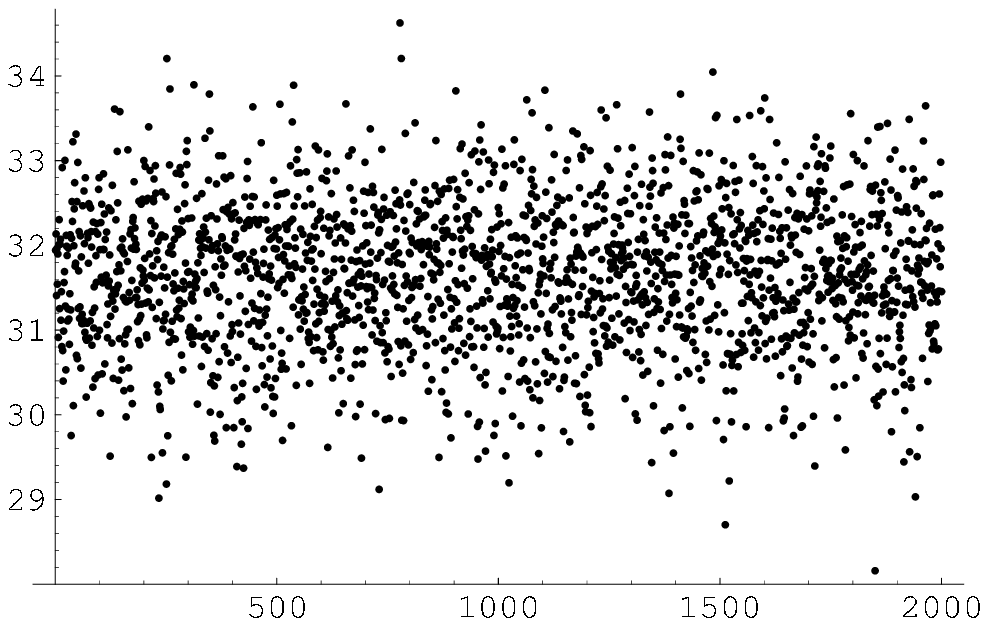}{8}{Typical radii for a configuration of individual particles (the data is by particle number)}

In this sense, in the large $N$ limit the numerical simulation of the density approaches a delta function, as the width of the distribution of particles in the radial direction goes down relative to the radius of the five-sphere.

\section{One hole states}

So far we have done a systematic analysis of the basic properties of the ground state wave functions for various $N$. Our results so far suggest that it is possible to work a detailed comparison between 
the numerical results of a Monte-Carlo analysis and the expectations that one has from a theoretical analysis of a formal $1/N$ expansion.

We would now like to analyze other configurations that in gravity could correspond to a non-trivial topology change. In particular, it is interesting to analyze the typical distribution of eigenvalues for the simplest LLM geometry whose topology is different than the ground state.

In particular, one can consider geometries that correspond to an annulus in the LLM plane. These are conjectured to be the result of condensing $H\sim N$ maximally giant gravitons associated to some particular half BPS orientation. A single such giant graviton is believed to  be given by the operator $\det Z$, and $H$ such gravitons would correspond to the operator $\det^H Z$. In the matrix model of eigenvalues, these would correspond to a wave function
\begin{equation}
\hat\psi_H \sim \hat \psi_0 \det(Z)^H
\end{equation}
and to an associated thermal ensemble
\begin{equation}
|\hat\psi|^2 = \exp( \sum (-\vec x^2+H\log((x^1)^2+(x^2)^2))+\sum\log(|\vec x_i-\vec x_j|)
\end{equation}
Thus the particles are repelled from the locus $x^1,x^2=(0,0)$. In two dimensions, such wave functions correspond to an inner circle of the annulus with a radius that scales with $\sqrt H$, independent of $N$.

Also, from the LLM intuition, one expects that in the LLM plane the region with eigenvalues (particles) is 
the degeneration locus of the $S^3$ of the boundary. At the edges of the droplets, both the $S^3$ of the boundary and the $S^3$ associated to $S^5$ being written as a fibration over a disc vanish. Thus we expect that if our particle distributions are to match the LLM topologies, then the projection of the particles to the $(1,2)$ plane will form an annulus, and at the edges, the size of the sphere in the orthogonal directions should vanish. As described previously, matching to an exact LLM metric is difficult, as it is not clear what is the coordinate change that relates the eigenvalue distributions to the LLM coordinates, and this map could be rather 
complicated. However, matching the topologies of the degeneration locus of the boundary $S^3$ should be straightforward.

\myfig{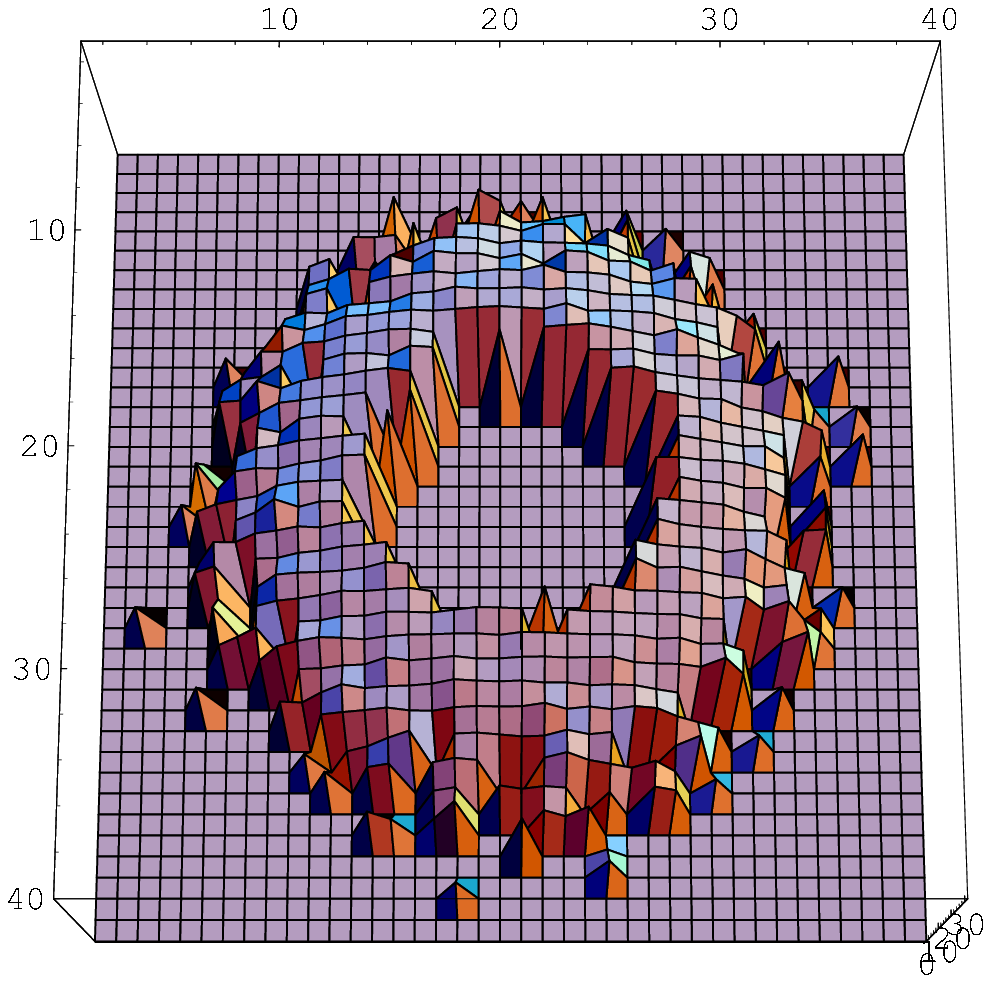}{7}{The plot shows the average radius of the particles inthe $3456$ directions
for binned data in the $12$ plane. The data shows a simulation with 2000 particles and $H=300$, in a $40\times 40$ grid.}

As can be seen from figure \ref{fig:avht.eps}, the distribution of particles projected on the plane forms an annulus shaped distribution of particles, exactly as expected from 
a naive analysis of the topology of such an LLM droplet.
In this sense, we are observing directly a topology change. However, the simulations do not show the expected closing of the distribution of particles into a donut, but rather there appears to be a boundary to the
surface geometry that the particles describe. We do not know at this moment if this is a feature of the physical system we are studying, or if we need to improve the code to focus on the region where the deviation from our naive intuition is taking place. This is currently
being investigated. For an alternative picture, we can plot the radius on the $12$ 
plane with respect to the radius in the $3456$ plane, as shown in figure \ref{fig:hsphere.eps}.

\myfig{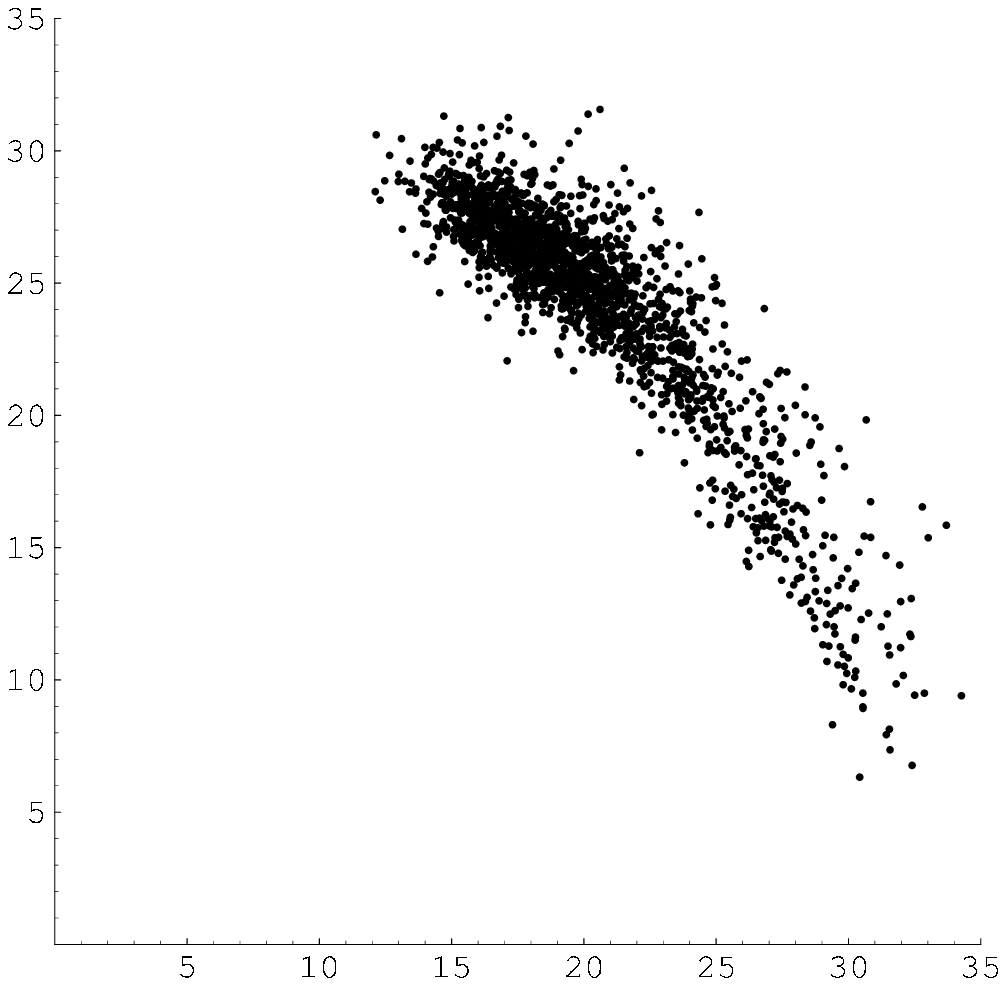}{5}{$r_{12}$ vs $r_{123456}$ for an $N=2000$ particle distribution, with
$H=30$. Notice the low number of points for $r_{12}$ large, relative to $r_{12}$ small: this is an effect of phase space for each type of configuration.}

If it is a new feature of the system, we need to explain the physics associated to it, and we may need to rethink some issues regarding what we mean by 
geometry associated to these droplets. We believe this is a very interesting result of the simulations we have done. Below, we will present evidence that this type of simulation is 
behaving like expected in terms of the $AdS$ intuition for the size of the hole in the 
center, giving us confidence that the distribution ending ``in the wrong place'' may be a physical feature that has to be considered seriously.
 
It is customary to 
compare LLM droplets directly with distributions of fermions associated to the lowest Landau
level in two dimensions for a given wave function. That is the simplest description of all half-BPS states \cite{Btoy, LLM}. Similarly, various studies of superstar geometries exploit this correspondence \cite{Buchel}. 

 For such wave functions, the size of the 
center hole in the Fermi-liquid picture is determined by the number of hole-particles that one places at the center of the ring, and this size is independent of the droplet. 

In our numerical simulations, we see a dependence of the size of the hole in terms of the total number of particles. This is shown in the table \ref{tab:holes}.  The way we determine the inner radius of the distribution is by averaging the five smallest values of $R_{12}$. The choice of $5$ is arbitrary and it is done to reduce large fluctuations of a single particle. The error in the simulations seems to be dominated by systematic errors, and we don't know how to estimate them, but we believe they can be bound by $\pm 1$ on all entries. Changing the coarse graining factor has an effect on the values obtained, and it is not clear how to compare different
distributions associated to the same number of particles $N_{eff}$, unless we put large systematic error bars of that size. Also, we have no good theoretical understanding of the fluctuations of the distributions. We can not compare results to a large $N$ limit saddle
point either, as we have no theoretical prediction of what the saddle point will look like. 

\begin{table}
\begin{center}
\begin{tabular} {|c|c|}
\hline N & $R_{in}$ \\ \hline
400 & $8.45\pm 1$\\
600 & $9.26\pm 1$\\
800 & $10.0\pm 1$\\
1200 & $11.1\pm 1$\\
1800 & $12.57\pm 1$\\
2400 & $13.34\pm1$\\
\hline
\end{tabular}
\caption{The above shows simulations with varying numbers of particles and fixed number of maximal giant gravitons ($H=30$). The right hand side is computed by averaging the five lowest values of $R_{12}$, the radius in the $12$ plane. } \label{tab:holes}
\end{center}
\end{table}

We can analyze the data In a log-log plot, as shown in figure{\ref{fig:loglog.eps} to test scaling, with $N$ at $H$ fixed. The data
can be fit linearly to very good accuracy 
to a slope of $0.26$, which seems very close to $1/4$.

\myfig{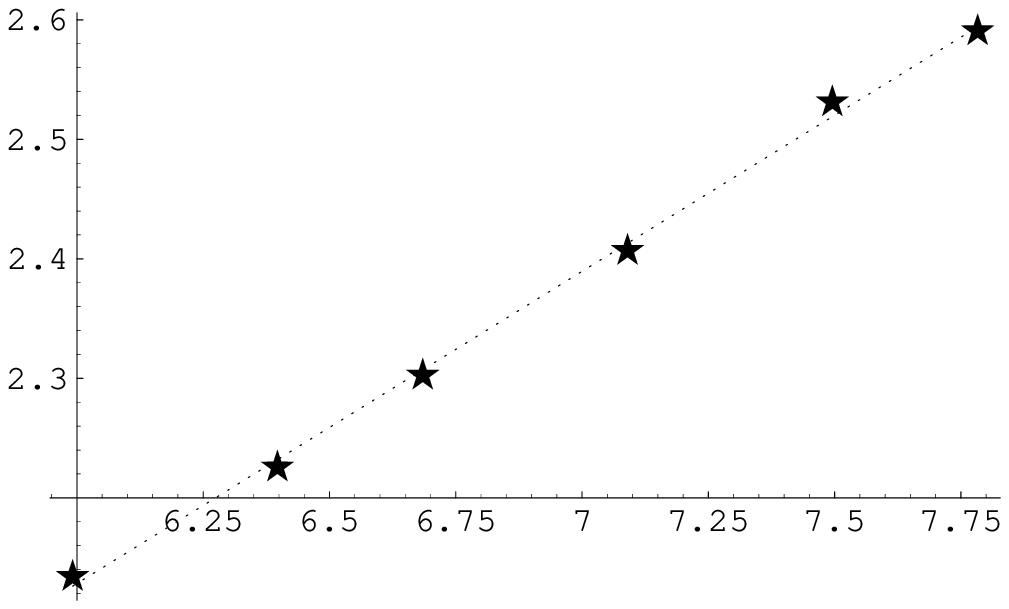}{5}{Log-Log data of size of central hole, as read from table 4.} }

Scaling of the saddle point suggests that the hole radius should be fixed relative to the 
size of the distribution if we keep $H/N$ fixed. In this fashion, we expect that $r_{in}
\sim g(H/N)$ is a function of $H/N$. For $H/N$ small, we seem to obtain simple scaling behavior.

 Thus, we expect that the radius of the hole is 
roughly given by $r\sim (HN)^{1/4}$, by guessing a simple rational exponent near the slope we found.

This result does not mean that the comparison with free fermion droplets are wrong. Rather, it shows that the LLM plane can be embedded very non-trivially into the $\BR^6$, with 
an as-yet unknown change of variables determining this embedding. It is also worth pointing out that the LLM metric is not associated to a flat metric on the LLM plane, but that there is also a warping by a
function $h^2$, as written in equation \ref{eq:LLM}. As discussed in section \ref{sec:AdS}, matching to an LLM geometry 
needs to resolve these changes of coordinates.

This factor of $N^{1/4}$ is reminiscent of the scaling of the string length. Indeed, we can motivate this comparison in a straightforward fashion. If we assume that we have $H$ D3-branes, where $H<<N$, we can think of the $H$ branes locally as flat D3-branes embedded in flat space, with small back reaction far away from the branes.  We would find then that the near geometry will start differing from a flat metric in a region of size $H^{1/4}$.

This is because for $H$ D3-branes, the metric of the geometry of the D3-branes in flat space is given  by \cite{HS}
\begin{equation}
ds^2 \sim f^{-1/2} dx_{||}^2 + f^{1/2}(dr^2+r^2 d\Omega_5^2)
\end{equation}
with 
\begin{equation}
f= 1+ \frac{4\pi  g_{YM}^2 H (\alpha')^2}{r^4}
\end{equation}
where $r$ is a radial direction transverse to the D-brane. The region where $f$ differs substantially from one is of order $r\sim H^{1/4}$, in string units. To be more precise, 
the region where we expect a discrepancy from flat space is of order $H^{1/4}$ in Planck units. Indeed, when we express $\alpha'$ in terms of $l_{pl}$, the factors of $g^2$ will cancel. This can be traced to the fact that the tension of the $D3$ brane does not depend on $g_s$ in units of $l_{pl}$.

Now, since in the dual $AdS$ geometry to the ${\cal N}=4 $ SYM with gauge group $U(N)$ we find that  $l_{pl}$ scales like
$N^{1/4}$, we find that having an inner radius that scales with $(HN)^{1/4}$ is exactly right
to match the intuition from $AdS$. We find this way that the matrix model we have written 
knows about the Planck scale in a way that can be measured geometrically.

This gives us evidence that the numerical results are capturing geometrical data 
associated to such a collection of $H$ D3-branes rather precisely, even if the comparison
to a particular LLM metric is hard.

\section{Conclusion and prospects}

In this paper we have explored how certain aspects of the $AdS/CFT$ correspondence can be studied numerically using Monte Carlo methods. We believe that we have shown that this is feasible: 
we can explore the large $N$ limit effectively with modest
computational resources. The main simplification that we have used is that at strong coupling  and for certain states only constant matrix fields contribute to the dynamics, and
moreover the dynamics reduces to configurations of commuting matrices.
Thus, for large matrices of rank $N$, the number of degrees of freedom that are relevant are of order $N$, as opposed to $N^2$. We also have a list of exact wave functions to simulate. Thus, we do not need to study the time evolution of $N$ degrees of freedom, but instead we can concentrate 
on thermodynamic properties of the degrees of freedom.

In particular, the most important description of the system under study is in terms of a distribution
of particles in six dimensions and the geometric properties of these distributions can be correlated with certain geometrical degenerations of the higher dimensional AdS dual 
geometry. 

To the extent that there are exact analytical results for these distribution, the numerical data seems to match the large $N$ limit very precisely. This is possible only for the ground state, where we have maximal rotational symmetry and is the only known distribution of particles in the large $N$ limit.

The distribution in the saddle point limit is a delta function. Since the radial fluctuations of the distribution are of order one, while the radius of the distribution is of order $N^{1/2}$, the 
distribution approaches a delta function in the limit.
It is expected that there might be $1/N$ corrections on the radius depending on what observables are chosen to define the radius itself. Many of these agree at large $N$, but differ by various $1/N$ corrections.

It is natural to imagine that these differences might account for various quantum gravity effects, as different probes will be sensitive to different definitions of the radius.

In some sense, the simulations we are doing encode all the quantum gravity corrections
to the geometry. However, the dictionary between our simulations and gravity needs to be improved.

We find that it is possible to consider other spacetime topologies. Guided by the intuition of the LLM metrics  \cite{LLM} and the construction of giant graviton state operators in the CFT \cite{BBNS}, we have been able to consider wave functions that give rise to measurable topology changes in the distributions of particles. In particular, we were able to show with these configurations that the
matrix model of six matrices is aware of the Planck scale size of the five sphere of
the dual $AdS_5\times S^5$ geometry, and therefore we have a measure of a 
relevant non-trivial scale in the problem.

It is also worthwhile to point out that the size of the sphere in field theory units is of order $\sqrt N$, 
and that each particle occupies a typical volume of size  $ N^{3/2}\sim (N^1/4)^5 N^{1/4}
\sim N^{1/4} l_{pl}^5$, much larger than a Planck scale volume. This shows that the AdS/CFT correspondence sees locality from a very different perspective than other discretizations of gravity, where the Planck scale is the scale of granularity of the discretization \cite{RidS}. Instead, we are able to see below the scale of discretization by studying collective effects and the Planck scale is related to studying the back reaction of the geometry to a heavy membrane. 
As a curiosity related to this problem, in figure \ref{fig:avht.eps} the particles accumulate densely in the region where geometry is behaving differently and it suggests that the density of resolution compensates for the scale of interesting features below the typical scale of granularity. 

We think it is important to emphasize that we are also very puzzled by the fact that the particle distributions are not behaving like we expected from a simple comparison to an LLM metric intuition.  This discrepancy might be due to  subtle issues regarding the simulation itself. On the other hand it is possible that we have not identified the precise correspondence to geometry in detail. Deciding between these options deserves further study.

We believe it is possible to make various measurements of correlators numerically and to compare them with the non-renormalization theorems for three point functions of BPS states.
This will require studying large samples of statistically independent configurations, necessitating an improved study of error analysis to complement the relatively crude estimates presented here.

It is straightforward to generalize our calculations to orbifold setups, following \cite{BC} and to more complicated LLM geometries. In particular, one can imagine that once it is better understood how to correlate the properties of the distributions to the metric aspects of the LLM geometries, it will be possible to study directly topology changing transitions, like the ones considered 
in \cite{HS}.

Another important aspect that might be studied with these methods is to consider a probe giant graviton that grows into the AdS region \cite{GMT, HHI}. Studying the corresponding wave functions in detail could give us a  precise coordinate map between the CFT field variables and the radial direction of the AdS geometry. 

One can also imagine using these numerical techniques to explore field theories that are not well understood, like the CFT duals associated to $AdS_4\times S^7$ or $AdS_7\times S^4$.
A recent proposal states that BPS states for the field theory duals are also described by a system of commuting matrices
\cite{BM}. Unlike the case of ${\cal N}=4 $ SYM, there is no measure that one can derive from first principles to study these objects as we have described in this paper. Numerical techniques might be very useful in making checks of a guess for that measure.

Finally, it would extremely interesting if one could extend these techniques further to study string states directly. This involves  introducing the off-diagonal modes systematically into this setup, as well as the other spherical harmonics of the quantum fields on the sphere \cite{B}. Some partial success has been achieved for adding strings to the geometric ground state in \cite{BCV}. However, one can study the same problem for other geometries where analytic string solutions are not available.

To conclude, we believe that the prospects for these types of simulations we have presented in this paper is full of open questions that can be addressed numerically with modest computer capabilities. It is our hope that numerical simulations might serve us a source of intuition 
for effects that we can not calculate otherwise.

\section*{Acknowledgements}

D.B. would like to thank S. Hartnoll, J. Liu, D. Stuart and R. Sugar for various discussions related to this project. The work of D. B. supported in part by  DOE, under grant DE-FG01-91ER40618.

\appendix

\section{Holomorphic wave functions}\label{app:hw}

We want to consider the following Hamiltonian for $N$ particles in $2d$ dimensions
\begin{equation}
H= \sum_i -\frac 1{2\mu^2} \nabla_i \mu^2 \nabla_i + \frac 12 |\vec x_i|^2 \label{eq:Aham}
\end{equation}
 where
\begin{equation}
\mu^2 = \prod_{i<j} |\vec x_i-\vec x_j|^2 \; .
\end{equation}
and wave functions of the form
\begin{equation}
\psi = \psi_0(X)  P(z) \label{eq:psin}
\end{equation}
where we follow the notation of section \ref{sec:MM}. In this notation $\psi_0$ is a gaussian eigenfunction of the Hamiltonian. 
\begin{equation}
\psi_0 \sim \exp( -\sum_i \vec x_i^2/2)
\end{equation} 
$P(z)$ is a homogeneous holomorphic polynomial on $3N$ complex variables 
\begin{equation}
z^s_j \sim  x_j^{2s-1}+ ix_j^{2s}.
\end{equation}  
The polynomial is invariant under permutation of the $\vec X_i$. The $z$ are holomorphic coordinates of the particles, after we have made a choice of complex structure on $\BR^{2d}$.
The Hamiltonian is the leading 
semiclassical approximation of the effective Hamiltonian that encodes all
BPS states associated to the chiral ring of ${\cal N}=4 $ SYM theory \cite{B}. We are interested in studying how good a description this Hamiltonian provides for the conjectured states that we have written in section \ref{sec:MM}. As we have seen, we have a ground state wave function $\psi_0$ that is gaussian. We want to show that the possible corrections to \ref{eq:psin}, or
to \ref{eq:Aham} are very small, and can be ignored.

When we calculate
\begin{equation}
\frac 1{\mu^2} \nabla_i \cdot \mu^2 \nabla_i \psi_0 P
\end{equation}
there are terms that contain no derivatives of $P$, terms with only one derivative acting on $P$, and terms that contain two derivatives of $P$. Let us separate these terms.

For the terms that do not depend on  derivatives of $P$, $P$ is acting just like a constant, and we can use the fact that $\psi_0$ is an eigenfunction of $H$ to show that these terms just give 
$E_0\psi_0$, the energy of the ground state. Now, the terms that contain two derivatives of $P$ are necessarily proportional to
\begin{equation}
(\nabla_i)^2 P =0
\end{equation}
and these vanish because $P$ is holomorphic. Indeed, is is possible to write 
\begin{equation}
\nabla_i^2 \sim \sum_s \partial_{z^s_i} \bar \partial_{\bar z_i^s}
\end{equation}

So we are only left to understand the terms that contain one derivative of $P$. These are given by
\begin{equation}
\frac 1{2 \mu^2} (\nabla_i P) \cdot\left( \psi_0 \nabla_i \mu^2 + \mu^2 2 \nabla_i \psi_0\right) 
\end{equation}
The second term gives us $\psi_0 x_i\cdot \nabla_i P$,  and when we sum over $i$,  we get the Euler vector for the scaling $\vec x_j\to \lambda x_j$. Since $P$ is a homogeneous polynomial of degree $m$, we find that $P$ is an eigenfunction of the corresponding scaling operator, with eigenvalue $m$.

The only obstruction for this wave-function to be an eigenfunction of the Hamiltonian is to check if
\begin{equation}
\sum_i \nabla_i P \cdot  \nabla_i \mu^2 
\end{equation}
vanishes or not. Now, we can easily show that
\begin{equation}
\nabla_i \mu^2 = \mu^2 \sum_{j\neq i} \frac{2(\vec x_i -\vec x_j)}{|\vec x^i -\vec x_j|^2}
\end{equation}
so that we can rewrite the expression in such a way that
\begin{equation}
\nabla_i P \nabla_i \mu^2 \sim  \sum_{i\neq j} \frac{1}{|\vec x_i-\vec x_j|^2}(\vec x_i-\vec x_j)\cdot (\nabla_i -\nabla_j) P 
\end{equation}
and there is no obvious mechanism for cancellations (except when $d=1$). 

What we want to do now is to think of this wave function as a variational approximation to an eigenfunction of the Hamiltonan. Seeing as different degree polynomials give orthogonal 
states (they have different $SO(6)$ quantum numbers), this is a well defined procedure to estimate the minimum energy solution of the Hamiltonian with given R-charge quantum numbers.

What we notice is that these terms that do not cancel can only be important when $|\vec x_i-\vec x_j|$ is small. For this situation we can ignore all other variables and concentrate on two of them ($\alpha$ and $\beta$ lets say), and expand $P$ in taylor series about their middle point $z^0$.

This is, we define
\begin{equation}
z^{\alpha} = z^0 +\delta z, z^\beta= z^0-\delta z
\end{equation}

If we define the following function of $z^0$ and $\delta z$ with all other particles at fixed position,
$f(z^0, \delta z)= P(z^{\alpha}, z^{\beta}) $,  then $f \sim P( z^0+\delta z, z^0 -\delta z) = P(z^0-\delta z, z^0+\delta z)$ because of the symmetry properties of $P$. From here 
we get 
$f(z^0, \delta z)= f(z^0,-\delta z)$. Also, from the change of variables we find that $\nabla_\alpha-\nabla_\beta \sim \nabla_{\delta z}$. We want to understand $f$ near the region where $\delta z=0$. From the symmetry properties, the first derivative of $f$ vanishes at $\delta z=0$, and we need to expand $f$ (the function $P$) to second order in
$\delta z$. We find then that the possible singular correction to the energy is finite (of order 
$f^{\prime\prime}/f$).

In this sense, the term that we can not control easily is bounded, and can be considered as a small perturbation near the region when the particles coincide. Elsewhere, 
the wave function is an approximate solution of the time independent Schrodinger equation.
For this last argument we have not exploited any additional properties of $P$ other than the symmetry under exchange of particles.

What we want to do now, is to be more careful about the second order expansion of what we have labeled $f$. Indeed, a Taylor expansion of $f$ shows us that
\begin{equation}
f(z^0, \delta z) \sim f(z_0,0) +\frac 12 \delta z^i \delta z^j f_{,ij}(z_0,0)+\dots
\end{equation}
where the partial derivatives in $f$ are taken with respect to the $\delta z$ variables.

We get then that
\begin{equation}
\frac {\partial f}{\partial \delta z^i} = f_{,ij}(z_0,0) \delta z^j
\end{equation}
and that we need to evaluate the average contribution to the energy by integrating this expression in the relevant region where this term is important, namely
\begin{equation}
 \int (\delta z)^d \frac{1}{|\delta z|^2}( \delta z^i) (\delta z^j) f_{,ij} 
\end{equation}
We notice that the integral over $\delta z$ on a small symmetric disc vanishes unless $i=j$ in the expression above, and that spherical symmetry guarantees that this is proportional to $f_{ii}(z_0,0)\sim \nabla_{\delta z}^2 f $.
Now we go back to our expression for $P$, and we remember that $P$ is holomorphic, so that 
$f$ is also a holomorphic function of $\delta z$ once we have chosen a complex structure for the $z^\alpha$. This means that for $P$ holomorphic, the term that is hard to compute averages to zero in the relevant region where it is important for the case of holomorphic functions.

This means that the energy of the state has to be given by 
\begin{equation}
E \sim E_0 + m
\end{equation}
where $m$ is the degree of $P$ and $E$ is extremely close to $E_0+m$. 

In the fully supersymmetric problem the corresponding BPS states are supposed to have energy given exactly by $E=E_0+m$, and all other states with the same $R$ charge have greater energy of order $1$. 
This implies that corrections from ignoring supersymmetry are small, and that we can assume
that in the worst case scenario, that the wave functions are not exact solutions of the supersymmetric problem, then the wave functions we have written are very good variational approximations to those wave functions.


\begin{thebibliography}{99}


\bibitem{RidS}
  D.~P.~Rideout and R.~D.~Sorkin,
  ``A classical sequential growth dynamics for causal sets,''
  Phys.\ Rev.\ D {\bf 61}, 024002 (2000)
  [arXiv:gr-qc/9904062].
  R.~D.~Sorkin,
  ``Causal sets: Discrete gravity,''
  arXiv:gr-qc/0309009.
  J.~Ambjorn, J.~Jurkiewicz and R.~Loll,
  ``Dynamically triangulating Lorentzian quantum gravity,''
  Nucl.\ Phys.\ B {\bf 610}, 347 (2001)
  [arXiv:hep-th/0105267].
  
\bibitem{M}
  J.~M.~Maldacena,
  ``The large N limit of superconformal field theories and supergravity,''
  Adv.\ Theor.\ Math.\ Phys.\  {\bf 2}, 231 (1998)
  [Int.\ J.\ Theor.\ Phys.\  {\bf 38}, 1113 (1999)]
  [arXiv:hep-th/9711200].


\bibitem{B}
  D.~Berenstein,
  ``Large N BPS states and emergent quantum gravity,''
  JHEP {\bf 0601}, 125 (2006)
  [arXiv:hep-th/0507203].

\bibitem{GKP}
  S.~S.~Gubser, I.~R.~Klebanov and A.~M.~Polyakov,
  ``Gauge theory correlators from non-critical string theory,''
  Phys.\ Lett.\ B {\bf 428}, 105 (1998)
  [arXiv:hep-th/9802109].

\bibitem{W}
  E.~Witten,
  ``Anti-de Sitter space and holography,''
  Adv.\ Theor.\ Math.\ Phys.\  {\bf 2}, 253 (1998)
  [arXiv:hep-th/9802150].
  
\bibitem{Btoy}
  D.~Berenstein,
  ``A toy model for the AdS/CFT correspondence,''
  JHEP {\bf 0407}, 018 (2004)
  [arXiv:hep-th/0403110].


\bibitem{HM}
  D.~M.~Hofman and J.~M.~Maldacena,
  ``Giant magnons,''
  J.\ Phys.\ A {\bf 39}, 13095 (2006)
  [arXiv:hep-th/0604135].
  
\bibitem{BCV}
  D.~Berenstein, D.~H.~Correa and S.~E.~Vazquez,
  ``All loop BMN state energies from matrices,''
  JHEP {\bf 0602}, 048 (2006)
  [arXiv:hep-th/0509015].


\bibitem{hat}
  Y.~Hatsuda and K.~Okamura,
 ``Emergent classical strings from matrix model,''
  arXiv:hep-th/0612269.

\bibitem{Bei}
  N.~Beisert,
  ``The su(2|2) dynamic S-matrix,''
  arXiv:hep-th/0511082.
  
\bibitem{SZ}
  A.~Santambrogio and D.~Zanon,
  ``Exact anomalous dimensions of N = 4 Yang-Mills operators with large R
  charge,''
  Phys.\ Lett.\ B {\bf 545}, 425 (2002)
  [arXiv:hep-th/0206079].

 
\bibitem{BFSS}
  T.~Banks, W.~Fischler, S.~H.~Shenker and L.~Susskind,
  ``M theory as a matrix model: A conjecture,''
  Phys.\ Rev.\ D {\bf 55}, 5112 (1997)
  [arXiv:hep-th/9610043].
  
\bibitem{BC}
  D.~Berenstein and D.~H.~Correa,
  ``Emergent geometry from q-deformations of N = 4 super Yang-Mills,''
  JHEP {\bf 0608}, 006 (2006)
  [arXiv:hep-th/0511104].
    D.~Berenstein and R.~Cotta,
  ``Aspects of emergent geometry in the AdS/CFT context,''
  Phys.\ Rev.\ D {\bf 74}, 026006 (2006)
  [arXiv:hep-th/0605220].
 
\bibitem{LLM}
  H.~Lin, O.~Lunin and J.~M.~Maldacena,
  ``Bubbling AdS space and 1/2 BPS geometries,''
  JHEP {\bf 0410}, 025 (2004)
  [arXiv:hep-th/0409174].
 
\bibitem{BIPZ}
  E.~Brezin, C.~Itzykson, G.~Parisi and J.~B.~Zuber,
  ``Planar Diagrams,''
  Commun.\ Math.\ Phys.\  {\bf 59}, 35 (1978).
 
\bibitem{Kleb}
  I.~R.~Klebanov,
  ``String Theory In Two-Dimensions,''
  arXiv:hep-th/9108019.

\bibitem{BBNS}
  V.~Balasubramanian, M.~Berkooz, A.~Naqvi and M.~J.~Strassler,
  ``Giant gravitons in conformal field theory,''
  JHEP {\bf 0204}, 034 (2002)
  [arXiv:hep-th/0107119].

\bibitem{McGST}
  J.~McGreevy, L.~Susskind and N.~Toumbas,
  ``Invasion of the giant gravitons from anti-de Sitter space,''
  JHEP {\bf 0006}, 008 (2000)
  [arXiv:hep-th/0003075].

\bibitem{GMT}
  M.~T.~Grisaru, R.~C.~Myers and O.~Tafjord,
  ``SUSY and Goliath,''
  JHEP {\bf 0008}, 040 (2000)
  [arXiv:hep-th/0008015].

\bibitem{HHI}
  A.~Hashimoto, S.~Hirano and N.~Itzhaki,
  ``Large branes in AdS and their field theory dual,''
  JHEP {\bf 0008}, 051 (2000)
  [arXiv:hep-th/0008016].
 
\bibitem{CJR}
  S.~Corley, A.~Jevicki and S.~Ramgoolam,
  ``Exact correlators of giant gravitons from dual N = 4 SYM theory,''
  Adv.\ Theor.\ Math.\ Phys.\  {\bf 5}, 809 (2002)
  [arXiv:hep-th/0111222].
 
\bibitem{BFHP}
  M.~Blau, J.~Figueroa-O'Farrill, C.~Hull and G.~Papadopoulos,
  JHEP {\bf 0201}, 047 (2002)
  [arXiv:hep-th/0110242].
  
\bibitem{BMN}
  D.~Berenstein, J.~M.~Maldacena and H.~Nastase,
  ``Strings in flat space and pp waves from N = 4 super Yang Mills,''
  JHEP {\bf 0204}, 013 (2002)
  [arXiv:hep-th/0202021].

\bibitem{HS}
  G.~T.~Horowitz and A.~Strominger,
  ``Black strings and P-branes,''
  Nucl.\ Phys.\ B {\bf 360}, 197 (1991).

\bibitem{GMNO}
 A.~Donos,
 ``A description of 1/4 BPS configurations in minimal type IIB SUGRA,''
  arXiv:hep-th/0606199.
  A.~Donos,
  ``BPS states in type IIB SUGRA with SO(4) x SO(2)(gauged) symmetry,''
  arXiv:hep-th/0610259.
  E.~Gava, G.~Milanesi, K.~S.~Narain and M.~O'Loughlin,
  ``1/8 BPS states in AdS/CFT,''
  arXiv:hep-th/0611065.

\bibitem{Kim}
  N.~Kim,
  ``AdS(3) solutions of IIB supergravity from D3-branes,''
  JHEP {\bf 0601}, 094 (2006)
  [arXiv:hep-th/0511029].


\bibitem{Chenetal}
B.~Chen et al.,
``Bubbling AdS and BPS states in IIB supergravity preserving an $S^3$ isometry'', {\em to appear}

\bibitem{Met}
N.~Metropolis, A.~W.~ Rosenbuth, M.~N.~Rosenbluth, A.~H.~Teller, E.~Teller, J. Chem. Phys.
21, 1087 (1953).

\bibitem{KT}
  D.~Kabat and W.~I.~Taylor,
  ``Spherical membranes in matrix theory,''
  Adv.\ Theor.\ Math.\ Phys.\  {\bf 2}, 181 (1998)
  [arXiv:hep-th/9711078].


\bibitem{LMRS}
  S.~M.~Lee, S.~Minwalla, M.~Rangamani and N.~Seiberg,
  ``Three-point functions of chiral operators in D = 4, N = 4 SYM at  large
  N,''
  Adv.\ Theor.\ Math.\ Phys.\  {\bf 2}, 697 (1998)
  [arXiv:hep-th/9806074].

\bibitem{DFS}
  E.~D'Hoker, D.~Z.~Freedman and W.~Skiba,
  ``Field theory tests for correlators in the AdS/CFT correspondence,''
  Phys.\ Rev.\ D {\bf 59}, 045008 (1999)
  [arXiv:hep-th/9807098].
  P.~S.~Howe, E.~Sokatchev and P.~C.~West,
  ``3-point functions in N = 4 Yang-Mills,''
  Phys.\ Lett.\ B {\bf 444}, 341 (1998)
  [arXiv:hep-th/9808162].

\bibitem{BCH}
D.~Berenstein, R.~ Cotta.
{\em Work in progress}



\bibitem{Buchel}
  M.~M.~Caldarelli and P.~J.~Silva,
  ``Giant gravitons in AdS/CFT. I: Matrix model and back reaction,''
  JHEP {\bf 0408}, 029 (2004)
  [arXiv:hep-th/0406096].
  A.~Buchel,
  ``Coarse-graining 1/2 BPS geometries of type IIB supergravity,''
  Int.\ J.\ Mod.\ Phys.\ A {\bf 21}, 3495 (2006)
  [arXiv:hep-th/0409271].
  N.~V.~Suryanarayana,
 ``Half-BPS giants, free fermions and microstates of superstars,''
  JHEP {\bf 0601}, 082 (2006)
  [arXiv:hep-th/0411145].
   M.~M.~Caldarelli, D.~Klemm and P.~J.~Silva,
 ``Chronology protection in anti-de Sitter,''
  Class.\ Quant.\ Grav.\  {\bf 22}, 3461 (2005)
  [arXiv:hep-th/0411203].
  G.~Mandal,
  ``Fermions from half-BPS supergravity,''
  JHEP {\bf 0508}, 052 (2005)
  [arXiv:hep-th/0502104].
  P.~G.~Shepard,
  ``Black hole statistics from holography,''
  JHEP {\bf 0510}, 072 (2005)
  [arXiv:hep-th/0507260].
  V.~Balasubramanian, J.~de Boer, V.~Jejjala and J.~Simon,
  ``The library of Babel: On the origin of gravitational thermodynamics,''
  JHEP {\bf 0512}, 006 (2005)
  [arXiv:hep-th/0508023].
   P.~J.~Silva,
  ``Rational foundation of GR in terms of statistical mechanic in the AdS/CFT
  framework,''
  JHEP {\bf 0511}, 012 (2005)
  [arXiv:hep-th/0508081].
  
\bibitem{Horava:2005pv}
  P.~Horava and P.~G.~Shepard,
  ``Topology changing transitions in bubbling geometries,''
  JHEP {\bf 0502}, 063 (2005)
  [arXiv:hep-th/0502127].
  
\bibitem{BM}
  S.~Bhattacharyya and S.~Minwalla,
  ``Supersymmetric States in M5/M2 CFTs,''
  arXiv:hep-th/0702069.
  
  
  
  
\end{thebibliography}
\end{document}